\documentclass[lettersize,journal]{IEEEtran}
\usepackage{amsmath,amsfonts,amssymb}
\usepackage{algorithmic}
\usepackage{algorithm}
\usepackage{array}
\usepackage[caption=false,font=normalsize,labelfont=sf,textfont=sf]{subfig}
\usepackage{textcomp}
\usepackage{stfloats}
\usepackage{url}
\usepackage{booktabs}
\usepackage{verbatim}
\usepackage{graphicx}
\usepackage{cite}
\usepackage{xurl}
\usepackage{orcidlink}
\usepackage{balance}
\usepackage{tikz}
\usetikzlibrary{positioning,fit,arrows.meta,backgrounds,calc,shadows}
\usepackage{amssymb}
\begin{document}

\title{SkillSelect-Serve: QoS-Aware Budgeted Skill Service Recommendation for LLM Agents}

\author{
Jingyuan Zheng,
Dongjing Wang*,~\IEEEmembership{Member,~IEEE,}
Xin Zhang,
Hao Chen,
Youhuizi Li,
Xudong Shen,
Haiping Zhang,
Butian Huang,
Dongjin~Yu,~\IEEEmembership{Senior Member,~IEEE,} and~Guandong~Xu,~\IEEEmembership{Senior Member,~IEEE}
\thanks{D. Wang* corresponding author, is with School of Computer Science and Technology, Hangzhou Dianzi University, Hangzhou, 310018, China and Sanya Traditional Chinese Medicine Hospital, Sanya, 572000, China. e-mail: Dongjing.Wang@hdu.edu.cn.}
\thanks{J. Zheng, X. Zhang, Y. Li, H. Zhang, and D. Yu are with School of Computer Science and Technology, Hangzhou Dianzi University. H. Chen is with the HDU-ITMO Joint Institute, Hangzhou Dianzi University. B. Huang is with the School of Cyberspace Security, Hangzhou Dianzi University.
X. Shen is with Netease Fuxi AI Lab, NetEase Inc, Hangzhou, China. G. Xu is with The Education University of Hong Kong, China. \textbf{E-mail}: \{zhengjoy, zhangxin, huizi, zhanghp, yudj, haoc, butian\}@hdu.edu.cn, hzshenxudong@corp.netease.com, gdxu@eduhk.hk}
}

\markboth{Manuscript submitted to IEEE Transactions on Services Computing}%
{Jingyuan Zheng \MakeLowercase{\textit{et al.}}: SkillSelect-Serve}


\maketitle

\begin{abstract}
Reusable agent skills are emerging as a service-oriented capability layer for Large Language Model (LLM) agents. Unlike plain retrieval items, a skill exposes functional capabilities, input-output assumptions, tool dependencies, context cost, and risk metadata. Selecting skills is particularly challenging for small LLM agents, which can load only a few capability units under restricted context, tool availability, and risk tolerance. Existing fixed Top-$k$ methods rank skills by textual relevance and overlook requirement satisfaction, deliverability, and operational constraints. To address these limitations, we present \textbf{SkillSelect-Serve}, a QoS-aware, budget-constrained Skill Service recommendation framework. Raw skills are profiled as structured Skill Services, and the task is converted into a structured requirement object. Candidates discovered from a large-scale Skill Service Registry are ranked by a calibrated task-conditioned suitability estimator and packed by a constrained projection enforcing token-budget, aggregated-risk, and tool-availability constraints, using only deployment-observable features. On a registry of 35,353 skills with pooled multi-positive relevance judgments verified by two independent assessors, the unconstrained top-5 recommendation fits a realistic 4,000-token context for only 9.1\% of tasks. The constrained projection restores 100\% deliverability at a cost of only 1.14 points of hit rate, outperforming retrieve-and-rerank, budget truncation, and diversity-based selection under identical budgets. The same mechanism halves delivered risk exposure and eliminates the 44--81\% tool-violation rates of tool-agnostic recommendation. At an identical three-service budget, hit rate improves from 0.8864 (fixed Top-3 retrieval) to 0.9091. Diagnostic execution studies further reveal a gap between offline recommendation quality and downstream small-agent execution. The results support managing reusable agent skills as discoverable, comparable, and constraint-aware service units instead of plain retrievable documents.
\end{abstract}


\section{Introduction}
\IEEEPARstart{L}{arge} Language Model (LLM) agents are evolving from single-prompt interactions into complex systems capable of task decomposition, tool invocation, and interaction with external APIs \cite{yao2022react,shen2023hugginggpt}. In this paradigm, agent skills serve as a critical intermediate layer between user tasks and executable capabilities: a skill packages natural-language instructions together with task workflows, tool-use specifications, code templates, input-output assumptions, execution constraints, and risk warnings. As such skills are accumulated, reused, and shared, large-scale skill libraries are giving rise to a new service-oriented ecosystem. Within this ecosystem, reusable skills act as service units that must be modeled, discovered, selected, and composed to support task execution by small LLM agents operating in constrained environments.

As a skill library scales to tens of thousands of entries, the core challenge is no longer to retrieve a semantically relevant skill, but to identify a \emph{set} of Skill Services that jointly satisfies the functional requirements of a task while remaining usable by the target agent. This problem is especially acute for small LLM agents, which can load only a few capability units under restricted context budgets, limited tool availability, and low risk thresholds: redundant services waste context, unavailable tools make relevant services unusable, and incompatible instructions introduce negative transfer. Effective recommendation must therefore consider not only textual relevance, but also service suitability, requirement coverage, compatibility, redundancy, context cost, and risk.

Existing methods for LLM tool use commonly cast external-capability integration as tool retrieval, API selection, routing, or reranking, scoring each candidate and returning a fixed number of top-$k$ items \cite{schick2023toolformer,qin2023toolllm,li2023api}. This paradigm has three structural limitations. First, it treats a skill as a retrievable document, neglecting the service attributes implicitly attached to it---input-output types, tool dependencies, context cost, risk level, and applicability conditions. Second, it treats each skill as the decision unit and therefore cannot capture complementarity, substitutability, or redundancy across skills. Two individually relevant skills may be largely redundant when combined, whereas a lower-ranked skill may supply a critical missing capability. Third, fixed top-$k$ selection assumes all tasks need the same number of skills, which cannot accommodate the budget constraints of small agents or support fine-grained trade-offs among coverage, cost, and risk.

The central argument of this paper is that reusable agent skills should be modeled and managed as service-oriented capability units rather than plain retrievable documents. We represent each skill as a Skill Service with functional capabilities, input-output assumptions, tool dependencies, context cost, and risk metadata---a view aligned with service computing research on service description, requirement-driven discovery, and QoS-aware selection \cite{papazoglou2007service,zeng2004qos,bouguettaya2017service}. The system then discovers candidates from a large-scale Skill Service Registry and recommends a feasible service bundle under the task requirements and the operational constraints of the target agent: a requirement-conditioned and budget-constrained service recommendation problem rather than document ranking.

Building on this formulation, we propose SkillSelect-Serve, a QoS-aware Skill Service recommendation framework for budget-constrained LLM agents. Raw skill documents are transformed into structured Skill Service Profiles exposing functional and non-functional attributes; a lightweight requirement planner converts the task into a structured requirement object without ever selecting skill identifiers. A shared discovery backbone retrieves a high-recall candidate set, over which dual-granularity modeling operates: a task-conditioned suitability estimator provides the calibrated, primary selection signal at the service level, while bundle-level calibrators estimate composition quality. A budget- and QoS-constrained projection finally packs the suitability ranking into the recommendation, skipping candidates that would violate service-budget, context-cost, tool-availability, or risk constraints, using only deployment-observable features. The framework thus improves selection \emph{within} a specified budget rather than loading more skills: under a strict budget it prioritizes a compact covering set, and under a larger budget it adds complementary services.

We evaluate SkillSelect-Serve on a Skill Service Registry of 35,353 deduplicated skills and 586 task requests, using pooled multi-positive relevance judgments on a held-out test set cross-checked by two independent assessors. Under a realistic 4,000-token context budget, the unconstrained recommendation is deliverable for only 9.1\% of tasks, whereas the constrained projection guarantees 100\% deliverability at a cost of 1.14 points of hit rate. The constrained projection also outperforms retrieve-and-rerank, budget-truncation, and diversity-based selection over the identical candidate pool and budget. The same mechanism reduces delivered risk exposure by 53\% at the constrained operating recall and eliminates the 44--81\% tool-violation rates incurred by tool-agnostic recommendations in restricted environments. These guarantees do not trade away selection quality: at an identical three-service budget, SkillSelect-Serve improves hit rate over fixed Top-3 retrieval from 0.8864 to 0.9091.

The main contributions of this paper are summarized as follows.
\begin{itemize}
    \item We formulate reusable agent skill selection as budget-constrained Skill Service recommendation: each skill is a service-oriented capability unit with functional, cost, and risk attributes, and deliverability is governed by three first-class QoS constraints---token budget, aggregated risk, and tool availability---connecting agent skill management with requirement-driven discovery and QoS-aware composition.

    \item We propose a requirement-conditioned, service-aware framework in which a calibrated task-conditioned suitability estimator drives selection, bundle-level calibrators estimate composition quality, and a constrained projection returns a bundle that is feasible by construction under the agent's operating envelope, exposing explicit recall--risk and recall--tool-availability frontiers.

    \item We build a rigorous evaluation protocol on a 35,353-skill registry---pooled multi-positive judgments in the TREC tradition verified by a second assessor, with strict train/development/test isolation---covering same-budget quality against strong selection baselines, granularity and feature ablations, upper-bound analysis, QoS feasibility frontiers, and a diagnostic execution study.
\end{itemize}

Our code and curated data are available at \url{https://github.com/ziyi0227/skillselect-serve}. The rest of this paper formulates the problem (Section~\ref{sec:problem}), details the SkillSelect-Serve framework (Section~\ref{sec:method}), describes the experimental setup (Section~\ref{sec:setup}) and results on five research questions (Section~\ref{sec:results}), discusses findings and limitations (Section~\ref{sec:discussion}), reviews related work (Section~\ref{sec:related}), and concludes (Section~\ref{sec:conclusion}).

\section{Problem Formulation}
\label{sec:problem}

This work focuses on recommending a feasible set of reusable agent skills from a large-scale registry for small LLM agent. Particularly, the recommendation decision must account for both functional requirements and operational constraints. We therefore model agent skills as Skill Services and formulate the task as requirement-conditioned and budget-constrained Skill Service recommendation.

\subsection{Skill Services, Tasks, and Agent Constraints}

Let the Skill Service Registry be $\mathcal{S}=\{s_1,\ldots,s_N\}$, where each skill is represented as a service-oriented capability unit: $s_i=\langle id_i,f_i,d_i,I_i,O_i,T_i,\Gamma_i,c_i,r_i \rangle .$
Here, $id_i$ is the service identifier; $f_i$ denotes the functional description and capability information; $d_i$ denotes the domain or task type; $I_i$ and $O_i$ denote the expected input and output types; $T_i$ denotes the required tools; $\Gamma_i$ denotes functional tags, dependencies, and other static metadata; $c_i$ denotes the context cost, measured by the number of tokens required to load the service; and $r_i$ denotes a metadata-derived risk attribute. This representation distinguishes a Skill Service from a plain text document: it exposes both functional properties (capabilities and interfaces) and non-functional properties (context cost and risk). Dynamic QoS attributes such as empirical reliability and latency can be incorporated when runtime logs are available, but are not required by the current formulation.

Given a natural-language task request $x$, the deployment environment of the target agent is represented as $A=\langle C_A,K_A,T_A,R_A\rangle$, where $C_A$ denotes the available context budget, $K_A$ the maximum number of Skill Services the agent can load, $T_A$ the tools accessible to the agent, and $R_A$ the acceptable risk threshold. For a particular operating point, the service budget $k$ satisfies $k\leq K_A$.

A lightweight requirement planner transforms the task into a structured service requirement object $\rho_x=\langle Cap_x,I_x,O_x,T_x,H_x,R_x\rangle$, where $Cap_x$ denotes the capabilities required by the task; $I_x$ and $O_x$ the task input and expected output; $T_x$ the tools potentially required; $H_x$ hard execution constraints, such as prohibited tools or mandatory output formats; and $R_x$ task-specific risk notes. The requirement object does not contain or generate service identifiers; it provides structured conditions for service discovery and recommendation.

\subsection{Discovery and Budget-Constrained Recommendation}

Because the registry may contain tens of thousands of Skill Services, directly evaluating every possible service bundle is infeasible. The system first performs requirement-conditioned service discovery, $\mathcal{C}_x=D(x,\rho_x,\mathcal{S})$ with $|\mathcal{C}_x|\ll|\mathcal{S}|$, whose goal is a high-recall candidate set rather than the final recommendation; its output combines textual retrieval signals with requirement-to-service matching signals such as capability, tool, and input-output matching.

For each candidate service $s_i\in\mathcal{C}_x$, the system estimates a task-conditioned service suitability score
\begin{equation}
p_i=f_{\mathrm{suit}}(x,\rho_x,s_i,A),
\label{eq:suit}
\end{equation}
which characterizes whether $s_i$ is suitable for the task and the target agent environment, conditioned on retrieval evidence, the Skill Service Profile, structured requirement matching, and operational compatibility.

Let $\mathcal{B}_x^{(k)}=\{B\mid B\subseteq\mathcal{C}_x,\ |B|\leq k\}$ denote the candidate bundle set for service budget $k$. In practice, $\mathcal{B}_x^{(k)}$ is a finite decision space constructed through ranking-based, coverage-oriented, diversity-oriented, and stochastic generation strategies rather than by enumerating all subsets of $\mathcal{C}_x$.

A defining property of the service setting is that the operational requirements of the target agent are not soft preferences to be traded off against relevance: a bundle that exceeds the context budget cannot be loaded, and a service whose required tools are unavailable cannot run. We therefore model deployability as a \emph{hard feasibility set} rather than as penalty terms in a composite objective:
\begin{equation}
\begin{aligned}
\mathcal{F}_x^{(k)}
=\{B\in\mathcal{B}_x^{(k)}\mid\
&\mathrm{Cost}(B)\leq C_A,\ \mathrm{Risk}(B)\leq R_A,\\
&\mathrm{Tools}(B)\subseteq T_A,\ \mathrm{HardSat}(\rho_x,B)=1\},
\end{aligned}
\end{equation}
where $\mathrm{Cost}(B)=\sum_{s_i\in B}c_i$ is the total context cost, $\mathrm{Risk}(B)$ is the aggregate risk exposure of the bundle, $\mathrm{Tools}(B)=\bigcup_{s_i\in B}T_i$ is its tool footprint, and $\mathrm{HardSat}(\rho_x,B)$ indicates whether the bundle satisfies the task-level hard constraints in $H_x$.

Within the feasible set, bundle quality is captured by a learned utility $U(x,\rho_x,B,A)$ that scores how well the bundle serves the task on the target agent. Its concrete realization---calibrated per-service suitability, optionally shaped by a risk penalty, together with bundle-level composition estimators---is instantiated in Section~\ref{sec:method}. The recommendation is the feasibility-constrained maximizer
\begin{equation}
B_k^\ast=\arg\max_{B\in\mathcal{F}_x^{(k)}} U(x,\rho_x,B,A).
\end{equation}

Two design choices deserve emphasis. First, the formulation contains no hand-weighted combination of relevance, redundancy, cost, and risk: requirements that admit no compensation---context capacity, tool availability, and task-level hard constraints---enter only through $\mathcal{F}_x^{(k)}$, while compensable preferences enter through the learned utility $U$. Second, risk deliberately appears in both places: as a hard ceiling $R_A$ for deployments that require an explicit guarantee, and optionally as a single tunable penalty inside $U$ that exposes a recall--risk operating frontier without weakening any guarantee. This separation mirrors the distinction between service-level agreements and QoS-aware preference optimization in classical service composition~\cite{zeng2004qos,zheng2010qos}.

\subsection{Problem Definition}

Given a task request $x$, a requirement object $\rho_x$, an agent environment $A$, a service budget $k$, and a registry $\mathcal{S}$, the objective is to discover a high-recall candidate set $\mathcal{C}_x$ and select from $\mathcal{B}_x^{(k)}$ the feasible bundle $B_k^\ast\in\mathcal{F}_x^{(k)}$ that maximizes $U(x,\rho_x,B,A)$. Feasibility subsumes the service-budget, context, tool, risk, and hard-constraint conditions; utility subsumes task-conditioned suitability, requirement coverage, complementarity, and redundancy control. This formulation differs from fixed top-$k$ retrieval by treating service attributes, bundle interactions, and agent-side operational constraints as explicit components of the recommendation problem.

\section{Method: SkillSelect-Serve}
\label{sec:method}

\begin{figure*}[t]
    \centering
    \resizebox{\textwidth}{!}{\input{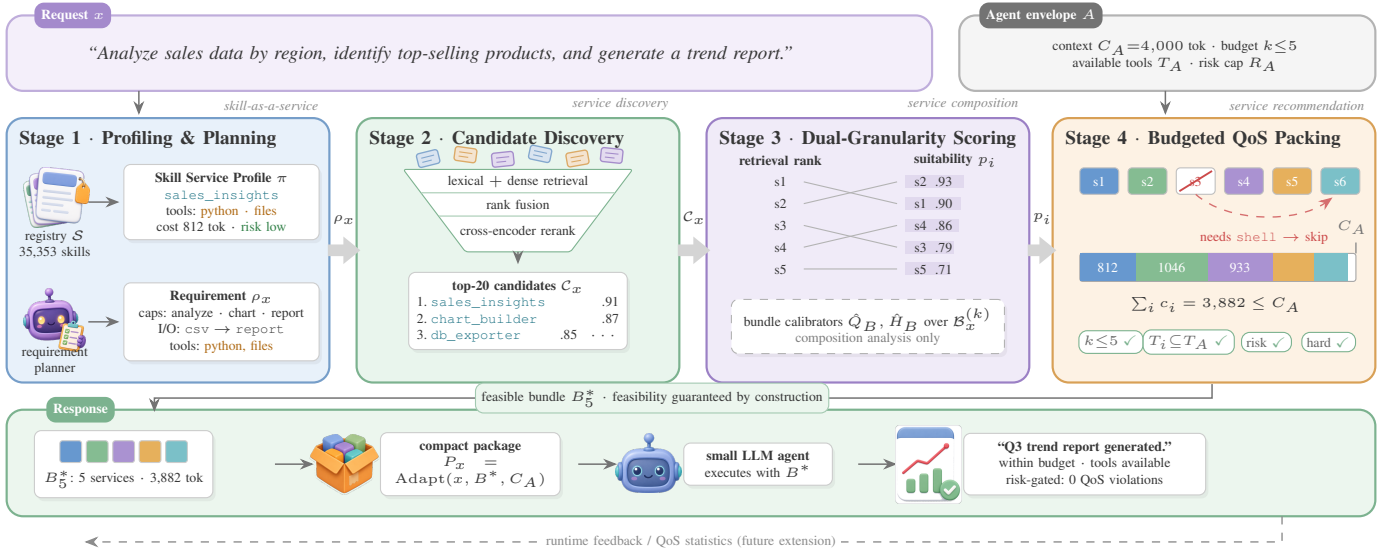}}
    \caption{Overview of SkillSelect-Serve. Raw skills are profiled as Skill Services; the requirement planner converts the task into a Structured Requirement Object that conditions candidate discovery; the recommendation core estimates task-conditioned service suitability and bundle-level composition quality; constrained selection returns a feasible bundle under service-budget, context-cost, tool-availability, risk, and task-level hard constraints.}
    \label{fig:overall_architecture}
\end{figure*}

SkillSelect-Serve implements the service-oriented decision chain defined above through four stages: Skill Service profiling, structured requirement planning and candidate discovery, service-aware dual-granularity modeling, and constrained bundle selection. As shown in Fig.~\ref{fig:overall_architecture}, given a task request $x$, an agent environment $A$, a service budget $k\leq K_A$, and a registry $\mathcal{S}$, the framework constructs structured Skill Service Profiles, converts the task into a requirement object $\rho_x$, retrieves a high-recall candidate set $\mathcal{C}_x$, estimates task-conditioned suitability at the service level and composition quality at the bundle level, and finally outputs the highest-scoring feasible bundle $B_k^\ast$ from $\mathcal{B}_x^{(k)}$. Retrieval rankings are never treated as final recommendations: retrieval only supplies the candidate space, and the subsequent modules decide suitability, composition quality, and deployment feasibility.

\subsection{Skill Service Profiling}
\label{sec:profiling}

SkillSelect-Serve first transforms raw skill documents into structured Skill Service Profiles. For each skill $s_i$, we construct the service description $\pi_i = \langle id_i, name_i, desc_i, domain_i, tools_i,$ $inputs_i, outputs_i, tags_i, cost_i, risk_i \rangle$, which instantiates the Skill Service representation introduced in Section~\ref{sec:problem}: the functional description $f_i$ is constructed from the skill name and textual description; $c_i$ is the token-based context cost; and $r_i$ is an LLM-annotated operational risk attribute.

This profiling step does not construct a complete formal service contract; it extracts lightweight service attributes from semi-structured skill documents, elevating a skill from an unconstrained text object to a discoverable, comparable, and constraint-aware service unit over which subsequent models can explicitly reason about tool dependencies, input-output compatibility, functional coverage, redundancy, cost, and risk.

In implementation, the service profile is constructed through metadata normalization, lightweight text analysis, and LLM-based attribute annotation. Identity, domain, tags, and input-output types are normalized from explicit metadata and document structure; context cost is measured with a calibrated characters-per-token accounting validated against production tokenizers. The two operational attributes that require semantic judgment---the risk level and the required-tool set---are annotated over the full registry by a cost-efficient LLM validated against a frontier reference: 95.8\% adjacent agreement on the four-level risk scale, versus 34.2\% for keyword heuristics, with disagreements concentrated in adjacent levels rather than in the high-risk decisions that drive gating (full validation protocol in the supplementary material). The resulting labels exhibit a plausible long-tail risk distribution with only 1.7\% high-risk skills, and are used as service-level features for discovery, suitability estimation, bundle construction, and constrained selection.

\subsection{Requirement Planner}

Natural-language task requests often imply multiple service requirements. If retrieval operates only on the raw query, the model may mistake textual similarity for service suitability; if an LLM is instead asked to select skill identifiers directly from a large registry, it may produce hallucinated or unreproducible choices that overlook QoS constraints. SkillSelect-Serve therefore introduces a lightweight requirement planner as a task-to-service requirement interface rather than a final selector: given a task request $x$, it outputs the structured requirement object $\rho_x=\langle Cap_x,I_x,O_x,T_x,H_x,R_x\rangle$ defined in Section~\ref{sec:problem} and never emits skill identifiers. The final bundle is selected downstream over the actual candidate set, which exploits the task-understanding ability of the small model while avoiding delegating discrete service selection to an LLM.

The requirement object is converted into explicit task-to-service matching features: capability matching between $Cap_x$ and $f_i$, input-output matching between $(I_x,O_x)$ and $(I_i,O_i)$, tool matching between $T_x$ and $T_i$, and hard-constraint satisfaction with respect to $H_x$. The risk notes $R_x$ describe task-level concerns, whereas $R_A$ is the maximum risk accepted by the target agent. The subsequent recommendation is therefore conditioned on service requirements rather than on raw textual similarity alone.

\subsection{Candidate Discovery and Bundle-Space Construction}

Following the formulation in Section~\ref{sec:problem}, the discovery backbone $D(x,\rho_x,\mathcal{S})$ produces a high-recall candidate set $\mathcal{C}_x$ by combining lexical retrieval, dense retrieval, rank fusion, and cross-encoder reranking. The retrieval ranking is used only for candidate generation: a highly ranked service may be redundant or infeasible in the final bundle, whereas a lower-ranked service may supply a missing capability, so the final decision is postponed to the selection stage. For each candidate service, the system retains its retrieval and reranking signals, Skill Service Profile features, requirement-matching features, context-cost features, and risk features.

Because enumerating all subsets of $\mathcal{C}_x$ is computationally infeasible, SkillSelect-Serve constructs the generated bundle space $\mathcal{B}_x^{(k)}$ with complementary strategies: prefix-based bundles preserve strong ranking baselines, top-combination bundles explore combinations among high-scoring candidates, coverage-oriented and low-redundancy bundles encourage requirement coverage and functional complementarity, and stochastic bundles increase compositional diversity. Each bundle is represented by a membership mask over $\mathcal{C}_x$ and characterized by bundle-level features such as aggregated suitability, requirement coverage, tool and input-output compatibility, domain diversity, selected-rank distribution, redundancy, token cost, and risk exposure. These bundles provide training instances for bundle-level calibration and an analyzable decision space for constrained selection.

\subsection{Service-Aware Dual-Granularity Modeling}

The core module of SkillSelect-Serve is service-aware dual-granularity modeling, which expands the recommendation core in Fig.~\ref{fig:overall_architecture} into two branches with an explicit division of labor. The \emph{service-level} branch---task-conditioned service suitability estimation---is the primary decision signal: it evaluates whether each candidate Skill Service matches the current task requirements and agent environment, and its calibrated scores directly drive the deployed constrained selection. The \emph{bundle-level} branch---quality and hit calibration---plays a complementary role: it estimates the overall usefulness of a service bundle and is used for composition-quality analysis and for the dual-granularity selector variants studied in the granularity ablation (Section~\ref{sec:ablation}). This primacy is an empirical finding rather than an assumption: as the ablation shows, selection driven by calibrated per-service suitability outperforms selection driven by bundle-level scoring alone, while the bundle-level signals remain informative for characterizing composition quality.

\subsubsection{Task-Conditioned Service Suitability}

Given a task request $x$, a structured requirement object $\rho_x$, an agent environment $A$, and a candidate service $s_i \in \mathcal{C}_x$, SkillSelect-Serve estimates the suitability score $p_i=f_{\mathrm{suit}}(x,\rho_x,s_i,A)$ defined in Section~\ref{sec:problem}. The score evaluates whether the service can support the current task under the target agent environment; it is not defined as the marginal utility of adding $s_i$ to an existing bundle. When the paper refers to \emph{marginal service suitability}, the term is used in the budgeted-selection sense---whether a service is worth one of the $k$ scarce slots---rather than as a bundle-conditional marginal gain.

The input to $f_{\mathrm{suit}}$ contains three groups of deployment-observable features. Discovery features include retrieval rank, lexical and dense scores, and reranker scores. Functional matching features include capability, tool, input, output, and hard-constraint matching. Operational features include context cost, risk, available-tool compatibility, and other agent-side constraints. During training, curated query-service labels provide supervision, and service suitability serves as a deployable enhancement to service discovery within the candidate set produced by the discovery backbone.

\subsubsection{Bundle-Level Quality and Hit Calibration}

Individual service suitability alone cannot characterize bundle-level interactions. Several highly suitable services may be functionally redundant, whereas services with moderate individual scores may jointly provide complementary capabilities. SkillSelect-Serve therefore reuses the bundle-level features $\phi(B,x,\rho_x,A)$ introduced during bundle-space construction and learns
\begin{equation}
\hat{Q}_B=f_{\mathrm{qual}}(\phi(B,x,\rho_x,A)),\qquad
\hat{H}_B=f_{\mathrm{hit}}(\phi(B,x,\rho_x,A)),
\end{equation}
where $\hat{Q}_B$ estimates the task-oriented quality of the bundle and $\hat{H}_B$ estimates the probability that the bundle contains at least one useful service. The quality calibrator captures bundle-level coverage, compatibility, complementarity, and interaction patterns that cannot be represented by individual service suitability alone. These calibrators are trained over the generated bundle space and serve as composition-quality estimators; they are not part of the deployed packing path (Section~\ref{sec:selection}). Context cost, risk, and tool availability enter the final selection stage as explicit feasibility constraints, so the paper distinguishes task-oriented bundle quality from deployment-time feasibility control.

Service-level supervision trains $f_{\mathrm{suit}}$, and bundle-level supervision trains $f_{\mathrm{qual}}$ and $f_{\mathrm{hit}}$; both are used only for training or candidate-space analysis. At inference, all decision inputs are deployment-observable---task requests, structured requirements, retrieval and reranking scores, Skill Service Profiles, requirement matching, compatibility, cost, risk, and redundancy---and exclude ground-truth labels, oracle utility, and label-derived symbolic utility. This separation is critical for deployable service recommendation, since real-world use cannot assume access to curated labels or oracle bundle utility.

\subsection{Budget- and QoS-Constrained Bundle Selection}
\label{sec:selection}

After obtaining calibrated service-level suitability scores, SkillSelect-Serve produces the final recommendation via \emph{constrained projection}: a greedy packing procedure that projects the suitability ranking into the feasibility set of Section~\ref{sec:problem}---guaranteeing feasibility by construction without claiming global optimality---with no hand-tuned trade-off weights.

The procedure operates directly on the suitability ranking. Candidates are ordered by the risk-shaped score $\tilde{p}_i = p_i - \lambda\, r_i$, where $\lambda\geq 0$ is the single optional risk-penalty knob and $\lambda=0$ recovers the pure suitability ordering. The projection then walks down this ordering with \emph{skip-and-continue} semantics: a candidate is admitted if and only if the bundle remains inside the feasibility set $\mathcal{F}_x^{(k)}$---its required tools are available ($T_i\subseteq T_A$), the accumulated context cost stays within $C_A$, the aggregated risk exposure stays within $R_A$, and the task-level hard constraints remain satisfied. Otherwise the candidate is skipped and the walk continues, so that budget freed by an infeasible candidate is reinvested in feasible substitutes further down the ranking rather than discarded. The walk terminates when $k$ services are selected or the candidate list is exhausted. Algorithm~\ref{alg:projection} summarizes the procedure.

\begin{algorithm}[t]
\caption{Constrained Projection with Skip-and-Continue Packing}
\label{alg:projection}
\begin{algorithmic}[1]
\REQUIRE candidates $\mathcal{C}_x$ with suitability $p_i$ and risk $r_i$; agent envelope $A=\langle C_A,K_A,T_A,R_A\rangle$; budget $k\leq K_A$; risk penalty $\lambda\geq 0$
\ENSURE feasible bundle $B$ with $|B|\leq k$
\STATE sort $\mathcal{C}_x$ by $\tilde{p}_i = p_i - \lambda r_i$ in descending order
\STATE $B\leftarrow\emptyset$;\ $cost\leftarrow 0$;\ $risk\leftarrow 0$
\FOR{each $s_i$ in sorted order}
  \IF{$|B|=k$} \STATE \textbf{break} \ENDIF
  \IF{$T_i\not\subseteq T_A$ \OR $cost+c_i>C_A$ \OR $risk+r_i>R_A$ \OR $\neg\mathrm{HardSat}(\rho_x,B\cup\{s_i\})$}
    \STATE \textbf{continue} \COMMENT{skip and reinvest the budget}
  \ENDIF
  \STATE $B\leftarrow B\cup\{s_i\}$;\ $cost\leftarrow cost+c_i$;\ $risk\leftarrow risk+r_i$
\ENDFOR
\RETURN $B$
\end{algorithmic}
\end{algorithm}

Two properties of this design matter for deployment. First, feasibility holds \emph{by construction}: every emitted bundle satisfies all hard constraints, so the 100\% feasibility reported in the experiments is a property of the algorithm rather than an empirical tendency. Second, skip-and-continue dominates naive truncation: truncation discards the tail of the ranking once the budget is exhausted, whereas the projection backfills feasible substitutes, spending the same budget on more services. The experiments quantify exactly how much recall this reinvestment recovers.

SkillSelect-Serve supports multiple operating points using the same learned estimators and shared constraint definitions, differing only in $k$: Compact@3 targets strict service budgets, Final@5 provides a moderate-budget default, and Aggressive@6 prioritizes broader task coverage when additional context is available. Compared with fixed top-$k$ retrieval, the selection stage does not follow the original retrieval ranking under the same budget but reselects services according to task-conditioned suitability and deployment constraints.

\subsection{Compact Skill Package for Diagnostic Execution}

For downstream diagnosis, the budgeted bundle can further be converted into a task-specific compact package $P_x = \mathrm{Adapt}(x, B^\ast, C_A)$, where $\mathrm{Adapt}(\cdot)$ keeps only the fragments most relevant to the current task---key execution steps, tool-use instructions, input-output constraints, and risk notes---under the context budget $C_A$. This module serves purely as a downstream execution interface for diagnosing the usability of the recommended services in small-model execution scenarios (Section~\ref{sec:results}). Runtime feedback for execution-aware QoS estimation is left as an extension toward a closed-loop service computing system.

\subsection{Complexity and Scalability}

The deployed decision path is lightweight. Skill Service profiling is a one-time offline pass over the registry, linear in the number of services $N$. At query time, discovery relies on standard lexical and dense retrieval indices, and reranking and suitability estimation each require one model evaluation per candidate, i.e., $O(|\mathcal{C}_x|)$ over a fixed-size pool ($|\mathcal{C}_x|=100$ in our experiments). The constrained projection sorts the candidates once and performs a single pass with constant-time budget, risk, and tool checks per candidate, so selection costs $O(|\mathcal{C}_x|\log|\mathcal{C}_x|)$ and is independent of registry size. The bundle-level calibrators operate only during training and analysis and add no cost to the deployed path. Per-query cost therefore grows with the candidate-pool size rather than with $N$, so the framework remains applicable to registries substantially larger than the 35,353-skill testbed.

\section{Experimental Setup}
\label{sec:setup}

\subsection{Skill Service Registry and Task Queries}

We evaluate SkillSelect-Serve on a large-scale Skill Service Registry containing 35,353 deduplicated skill items, each transformed into a structured Skill Service Profile with functional descriptions, tool dependencies, input-output types, functional tags, token cost, and risk level (Section~\ref{sec:profiling}). The tool vocabulary contains 12 canonical tools, and 56.9\% of the skills declare no tool dependency; only 1.7\% of the skills are labeled high-risk, which reflects the long-tail nature of operational risk in real skill ecosystems.

The task set contains 586 natural-language task queries, of which 577 have curated positive skill interactions and are evaluable. We partition the evaluable queries into 403 training, 86 development, and 88 test queries. The curated interactions (717 positives in total, 1.24 per query on average) are used \emph{only} to construct training supervision and development-time model selection; no test query is exposed to any learned component during training. All headline results in this paper are reported on the 88 held-out test queries.

Unlike conventional tool-use benchmarks that focus on the correctness of API calls~\cite{qin2023toolllm,li2023api}, this paper focuses on budgeted service recommendation and composition in a large-scale skill service ecosystem. Since the focus is the selection and composition stage, all compared methods use the same candidate discovery backbone, ensuring that differences among methods arise from service selection decisions rather than from differences in the candidate pool.

\subsection{Multi-Positive Relevance Judgments}
\label{sec:multigold}

A large skill registry inevitably contains multiple functionally interchangeable services for the same task; a single curated positive per query would therefore systematically under-credit any method that selects a valid alternative. To obtain a faithful evaluation signal, we construct pooled multi-positive relevance judgments for the test set, following the standard TREC pooling methodology~\cite{voorhees2005trec}. For each of the 88 test queries, we pool the top-20 candidates produced by the system runs and grade each pooled skill on a three-level scale (0: not relevant; 1: partially relevant; 2: directly usable). The primary assessor is a frontier LLM (Claude Sonnet), and the original curated positives are anchored as directly usable. The resulting \emph{core} positive set (grade-2 skills together with the curated anchors) contains 5.14 positives per query on average, and 88.6\% of the test queries have more than one core positive, confirming that the single-positive assumption would substantially distort evaluation.

To guard against single-assessor bias, a second independent LLM assessor (DeepSeek V4-Flash) re-graded the full pool. Inter-assessor agreement reaches Cohen's $\kappa=0.62$ on the three-level scale and $\kappa=0.70$ on the binary core decision, both in the \emph{substantial} range and comparable to or higher than typical human inter-assessor agreement reported in TREC settings. All headline recall numbers in this paper are computed against the primary core set; we verified that the conclusions are stable under the second assessor's judgments and under the intersection of both (Section~\ref{sec:qos}).

\subsection{Compared Methods}

We compare five categories of methods. \emph{(i) Fixed-budget prefix baselines.} Top-1, Top-3, and Top-5 Fixed select the highest-ranked candidates according to the calibrated relevance ranking, representing the conventional top-$k$ retrieval or routing paradigm.

\emph{(ii) Budget-aware selection baselines} isolate the contribution of the selection strategy: all consume the \emph{same} top-100 candidate pool, per-skill token accounting, and budget configuration (4{,}000 tokens, $k{=}5$), differing only in bundle selection. CE-Rerank Top-$k$ takes the budget-feasible prefix of a cross-encoder reranking, representing recent skill-routing systems; Relevance Top-$k$ takes the budget-feasible prefix of the calibrated marginal ranking; Greedy Knapsack packs skills by relevance-per-token density, the strongest classical budgeted heuristic; and MMR~\cite{carbonell1998mmr} with $\lambda\in\{0.7, 0.85\}$ is the textbook diversity-aware strategy.

\emph{(iii) Heuristic VarSize} adaptively determines the bundle size from ranking scores, redundancy, and thresholding rules---a strong manually designed variable-size baseline that learns no service utility.

\emph{(iv) Learning-based ablation variants.} Pure Neural Bundle Composer scores candidate bundles with a neural bundle scorer only; Dual Bundle Calibrator estimates utility and hit likelihood at the bundle level; the Skill-level Marginal Calibrator estimates per-service marginal suitability and is a key component of the deployed system.

\emph{(v) SkillSelect-Serve operating regimes.} Compact@3, Final@5, and Aggressive@6 instantiate the same budgeted QoS-aware projection under $k=3$, $5$, and $6$; the QoS-constrained mode additionally enforces the token budget with skip-and-continue packing and applies risk and tool-availability gates. In addition, we report the candidate-space upper bound---the recall attainable by an oracle selector over the candidate pool---which is not a deployable method but locates whether residual error lies in discovery or in selection.

\subsection{Evaluation Metrics}

We evaluate the compared methods from four perspectives: recommendation quality, composition quality, QoS feasibility, and execution behavior. Unless otherwise stated, all metrics are computed on the 88 held-out test queries against the core multi-positive judgments (Section~\ref{sec:multigold}). The deployed pipeline is deterministic at inference time---frozen retrievers, fixed calibrators, and a deterministic constrained projection---so repeated runs yield identical recommendations; we therefore report exact metric values rather than means and standard deviations over random seeds.

\textbf{Hit Rate} is the primary recommendation metric: the fraction of queries whose recommended bundle contains at least one core-relevant skill service. It reflects whether the system can surface a critical service under a limited budget, and is robust to the presence of interchangeable positives.

\textbf{Coverage Recall} measures composition quality under multiple positives. For a query with core set $G$ and bundle $B$ of size budget $K$, it is defined as $|B \cap G| / \min(K, |G|)$. Whereas hit rate credits a single correct pick, coverage recall rewards bundles that assemble complementary relevant services, and penalizes selections that spend the budget on redundant or irrelevant items.

\textbf{Bundle Size and Token Footprint} denote the number of recommended services and their total context cost in tokens. Token cost is measured with a characters-per-token accounting calibrated against production tokenizers; all budget-related conclusions are additionally verified with exact tokenizer counts (Section~\ref{sec:qos}).

\textbf{QoS Feasibility Metrics} include the budget feasibility rate (fraction of queries whose bundle fits the token budget), the delivered risk exposure (sum of the risk scores of the recommended services, with risk levels none/low/medium/high mapped to $0/0.25/0.55/1.0$), and the tool violation rate (fraction of bundles containing a service whose required tools are unavailable in the deployment environment). These metrics quantify whether a recommendation is actually deliverable to a constrained agent, not merely relevant.

\textbf{Ranking Metrics} for component-level analysis include Recall@$k$ and average precision (AP) of the marginal suitability ranking, which we use in the feature ablation study.

In the diagnostic execution study, we further report pass rate and average judge score. This experiment is intended to diagnose whether the recommended skill package can assist small-agent execution, rather than to serve as the main state-of-the-art claim of this paper.

\subsection{Implementation Details}

The candidate discovery backbone combines lexical retrieval with BM25~\cite{robertson2009probabilistic}, dense semantic matching with a frozen all-MiniLM-L6-v2 sentence encoder~\cite{reimers2019sentence}, and reciprocal rank fusion~\cite{cormack2009reciprocal}. A cross-encoder reranker initialized from ms-marco-MiniLM-L-6-v2~\cite{nogueira2019passage}, implemented in PyTorch with the Sentence-Transformers library, is fine-tuned strictly on the 403 training queries; the frozen first-stage retrievers contain no learned parameters.
Fine-tuning runs for two epochs with the AdamW optimizer at a learning rate of $2\times10^{-5}$, a batch size of 16, a maximum sequence length of 512 tokens, and linear warmup over the first 10\% of optimization steps, on a single NVIDIA RTX 5090 GPU. This split discipline guarantees that no test query influences any trained component. All selection methods consume the same candidate set, which controls for discovery differences and keeps the experimental focus on budgeted service selection.

The marginal service suitability estimator takes query-skill-level label-free features as input, including retrieval rank, reranker score, service profile features, requirement-matching features, tool and input-output compatibility, cost, and risk. The bundle-level calibrators estimate bundle utility and hit likelihood from features of candidate bundles, such as aggregated marginal suitability, coverage, redundancy, cost, risk, and rank distribution. The calibrators adopt histogram-based gradient-boosting models~\cite{friedman2001greedy} implemented in scikit-learn; on the held-out pair-level split used during calibrator development, the marginal calibrator reaches an AP of 0.851 for individual-service relevance (the layered ablation in Table~\ref{tab:ablation_granularity} retrains per feature layer on the official test partition, hence a different AP scale).

The default QoS-constrained configuration uses a token budget of 4{,}000, service budget as $k=5$, skip-and-continue packing (Section~\ref{sec:selection}), an aggregated risk budget, and tool-availability gate parameterized by deployment environment.\footnote{To support reproducibility, we will release the preprocessing scripts, service-profile construction code, evaluation scripts, and anonymized task-level prediction outputs upon publication, after removing private identifiers, unsafe execution traces, and content subject to redistribution constraints.}

\section{Experimental Results}
\label{sec:results}

This section answers five research questions. \textbf{RQ1} asks whether recommendations can be made \emph{deliverable} to a small agent under the three QoS constraints that govern deployment (token budget, risk exposure and tool availability) and whether the constrained projection spends the budget more effectively than standard selection strategies. \textbf{RQ2} examines whether the underlying budgeted selection improves recommendation quality against fixed top-$k$ retrieval under the same service budget. \textbf{RQ3} analyzes which modeling granularity and feature groups drive the selection signal. \textbf{RQ4} locates the system bottleneck through candidate-space upper-bound analysis. RQ5 conducts a small-agent execution diagnostic relating offline recommendation quality to actual execution utility.

\subsection{RQ1: Can Recommendations Be Made Deliverable under QoS Constraints?}
\label{sec:qos}

A recommendation that does not fit the target agent's operating envelope is not a service, regardless of its relevance. We therefore begin with the question that most sharply separates skill-service recommendation from document ranking: whether the recommended bundle can actually be delivered to a small agent under its token budget, risk tolerance, and available tools. All results are on the 88 test queries with core multi-positive judgments.

\begin{figure*}[t]
    \centering
    \includegraphics[width=0.92\textwidth]{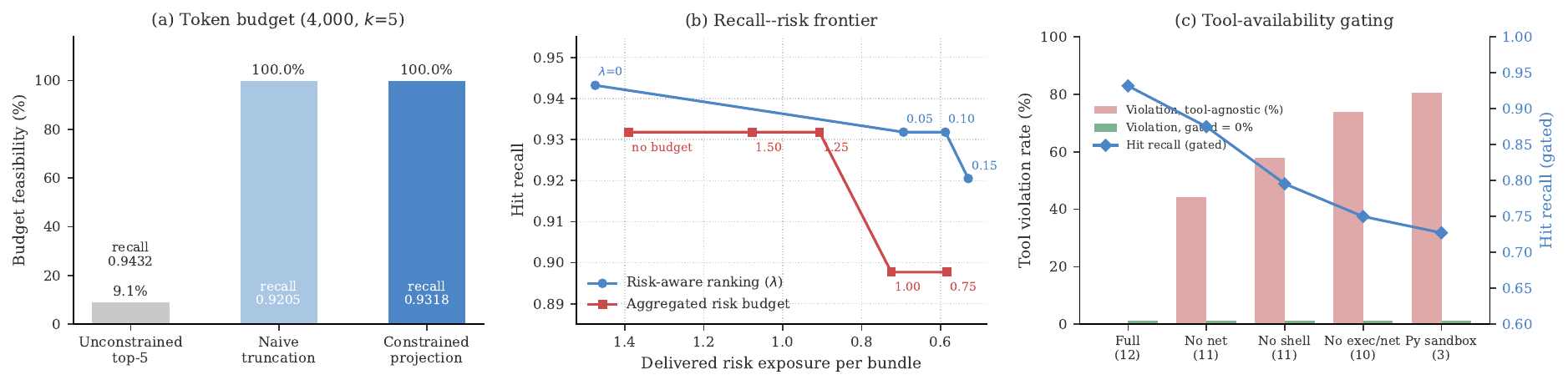}
    \caption{QoS operating frontier (88 held-out test queries). (a) Token budget: the unconstrained top-5 recommendation fits a 4{,}000-token context for only 9.1\% of queries; naive truncation and the constrained projection both restore 100\% feasibility, but the projection retains a higher hit rate (0.9318 versus 0.9205). (b) Recall--risk frontier: risk-aware ranking ($\lambda$) and aggregated risk budgets trace complementary paths; delivered exposure drops by 53--60\% at essentially unchanged recall. (c) Tool-availability gating across five deployment environments: tool-agnostic recommendations violate tool availability on 44--81\% of queries, the gate restores 0\% violations, and the recall cost is monotone in environment scarcity.}
    \label{fig:qos_frontier}
\end{figure*}

\textbf{Token budget: from 9\% feasible to 100\% feasible at 1.14 points.} Table~\ref{tab:selection_baselines} and Fig.~\ref{fig:qos_frontier}(a) examine the load-bearing constraint. The unconstrained top-5 recommendation averages 4{,}882 tokens against a 4{,}000-token context budget and fits the budget for only 9.1\% of queries: for a small agent, the unconstrained recommendation is undeliverable nine times out of ten. Naive truncation (the budget-truncated relevance prefix in Table~\ref{tab:selection_baselines}) restores feasibility but discards the tail of the ranking, wasting 12\% of the budget and dropping hit rate to 0.9205. The constrained projection with skip-and-continue packing also guarantees 100\% feasibility, yet keeps 4.53 services and 0.9318 hit rate---a total cost of 1.14 points relative to the infeasible ceiling. Because these conclusions could hinge on the token accounting, we re-ran the entire pipeline with exact tokenizer counts (o200k vocabulary). The character-based accounting proves conservative (actual 4.28 characters per token), naive feasibility is 17.05\% instead of 9.1\%, and the feasibility and recall conclusions are unchanged.

\textbf{Head-to-head against strong selection strategies.} Guaranteed feasibility alone would mean little if the budget were spent poorly. Table~\ref{tab:selection_baselines} therefore isolates the selection strategy under the same deployment regime: every method consumes the same top-100 calibrated candidate pool,  per-skill token accounting,  constraints (token budget 4{,}000, $k=5$), and differs only in how it selects the bundle.

\begin{table*}[t]
\centering
\caption{Budget-constrained selection head-to-head (88 test queries; identical candidate pool, token accounting, and budget: 4{,}000 tokens, $k{=}5$). Coverage recall is normalized by $\min(5,|G|)$. Feasibility is the fraction of queries whose bundle fits the token budget; the exact-tokenizer figure (o200k vocabulary) is in parentheses. The unconstrained ceiling ignores the token budget.}
\label{tab:selection_baselines}
\begin{tabular}{lcccccc}
\toprule
Selection Strategy & Hit Rate & Coverage Recall & Relevant / Query & Avg. Size & Avg. Tokens & Feasible \\
\midrule
CE-Rerank Top-$k$ (retrieve-and-rerank) & 0.8977 & 0.5797 & 2.09 & 3.66 & 3401 & 100\% \\
Relevance Top-$k$ (budget-truncated prefix) & 0.9205 & 0.5655 & 2.01 & 3.52 & 3342 & 100\% \\
Greedy Knapsack (density packing) & 0.9205 & 0.5987 & 2.17 & 4.65 & 3718 & 100\% \\
MMR ($\lambda=0.7$) & 0.9205 & 0.3602 & 1.11 & 4.56 & 3785 & 100\% \\
MMR ($\lambda=0.85$) & 0.9205 & 0.3896 & 1.25 & 4.61 & 3801 & 100\% \\
SkillSelect-Serve (constrained) & \textbf{0.9318} & \textbf{0.6053} & \textbf{2.18} & 4.53 & 3816 & \textbf{100\%} \\
\midrule
Unconstrained ceiling (top-5, no budget) & 0.9432 & 0.6621 & 2.47 & 5.00 & 4882 & 9.1\% (17.1\%) \\
\bottomrule
\end{tabular}
\end{table*}

Three observations follow. First, the learned marginal calibration matters: even the plain budget-truncated prefix of the calibrated ranking outperforms the cross-encoder retrieve-and-rerank strategy by 2.28 points of hit rate (0.9205 versus 0.8977), and the full constrained selector extends the margin to 3.41 points of hit rate and 2.56 points of coverage recall. Second, budget-aware packing matters: skip-and-continue packing fills the budget with 4.53 services instead of truncating at 3.52, improving hit rate by 1.13 points and coverage recall by 3.98 points over naive truncation. Third, diversity-oriented selection is actively harmful in this regime: MMR loses 21.6--24.5 points of coverage recall because penalizing similarity evicts relevant services that are legitimately close to each other in a registry full of near-duplicate capabilities.

The strongest classical heuristic, greedy knapsack packing, comes close to SkillSelect-Serve (0.9205 versus 0.9318 hit rate; 0.5987 versus 0.6053 coverage recall). On relevance metrics alone, density-based packing is a competitive selection rule. The advantage of the constrained-projection formulation is that it treats the token budget as one instance of a general QoS constraint family, so risk budgets and tool-availability gates integrate into the same selection mechanism without ad-hoc modification, as the remainder of this section demonstrates. The greedy density rule, by contrast, has no principled extension to non-additive feasibility constraints.

\textbf{Robustness across assessors.} Under the second assessor's judgments the constrained projection scores 0.9545 against a 0.9659 unconstrained ceiling, and the primary numbers equal those computed on the intersection of both assessors' positive sets, so the deliverability conclusions are not artifacts of a single assessor.

\textbf{Risk: a compressible QoS dimension.} With LLM-annotated risk labels, only 1.7\% of registry services are high-risk, but medium-risk services are common: the unconstrained top-5 bundle delivers an aggregated risk exposure of 1.475 per query (risk levels is $0/0.25/0.55/1.0$). Two complementary mechanisms operate, tracing the frontier in Fig.~\ref{fig:qos_frontier}(b) (full trade-off table in the supplementary material). First, \emph{risk-aware ranking} subtracts a risk penalty $\lambda$ from the calibrated score before selection: at $\lambda=0.05$ the delivered exposure drops by 53\% (1.475 to 0.694) while hit rate moves from 0.9432 to 0.9318---the same level as the budget-constrained operating point---and at $\lambda=0.1$ exposure drops by 60\% at unchanged recall. Relevant services usually have low-risk substitutes in a large registry, so risk is a highly compressible dimension. Second, for deployments that require a hard guarantee, an \emph{aggregated risk budget} in the constrained projection enforces a per-bundle exposure ceiling: tightening the budget to 1.5 and 1.25 leaves hit rate at 0.9318 while cutting exposure by 23\% and 35\%, even though 49--58\% of unconstrained bundles would have violated these ceilings. Tightening further to 1.0 trades 3.4 points of recall for a 48\% exposure reduction. The system thus exposes a recall--risk Pareto frontier with explicit service-level guarantees in the spirit of aggregated QoS-aware service composition~\cite{zeng2004qos,zheng2010qos}.

\textbf{Tool availability: a hard feasibility constraint.} Unlike risk, a missing tool cannot be compensated: a skill that requires shell access simply cannot run in a sandbox without it. We evaluate tool-availability gating across five deployment environments defined over the 12-tool vocabulary (Fig.~\ref{fig:qos_frontier}(c); full per-environment table in the supplementary material). The problem is real: in restricted environments, 44--81\% of tool-agnostic recommendations contain at least one unrunnable service. The tool-availability gate restores a 0\% violation rate in every environment while keeping bundle sizes essentially intact (4.3--4.5 services), by backfilling feasible substitutes rather than merely deleting offenders. The recall cost degrades gracefully with environment scarcity: relative to the full-toolset reference of 0.9318, air-gapped deployment costs 5.7 points, whereas a three-tool Python sandbox costs 20.5 points---the price of guaranteed runnability, which quantifies how much of the registry's value is locked behind each tool.

Together, the three dimensions form a coherent QoS profile. The token budget is the load-bearing constraint that determines whether a recommendation is deliverable at all; risk is a soft, highly compressible dimension where near-free reductions are available; and tool availability is a hard environmental constraint whose price is explicit and monotone in environment scarcity. Because all three enter the same constrained projection, an operator can configure the operating point per deployment---strict sandbox, air-gapped, or full-privilege---rather than accepting a single fixed recommendation policy.

\smallskip
\noindent\fbox{\parbox{0.97\linewidth}{\textbf{Findings (RQ1):} The constrained projection makes recommendations deliverable by construction: token-budget feasibility rises from 9.1\% to 100\% at a 1.14-point hit-rate cost, delivered risk exposure is halved at essentially unchanged recall, and the 44--81\% tool-violation rates are eliminated---while outperforming retrieve-and-rerank, budget truncation, and diversity-based selection under identical budgets.}}

\subsection{RQ2: Does Budgeted Selection Improve Recommendation Quality at Equal Budget?}
\label{sec:quality}

The deliverability guarantees of RQ1 would mean little if they were built on weak selection. RQ2 therefore examines the recommendation quality of the underlying budgeted selection when the budget is expressed as a service count, against the fixed top-$k$ paradigm that dominates current skill-selection practice. Table~\ref{tab:main_results} reports the main recommendation results on the 88 held-out test queries under the core multi-positive judgments.

\begin{table}[t]
\centering
\caption{Results on budgeted skill-service recommendation (88 held-out test queries, core multi-positive judgments). Coverage recall is normalized by $\min(\text{size},|G|)$, favoring smaller bundles for adaptive-size methods.}
\label{tab:main_results}
\footnotesize
\setlength{\tabcolsep}{3pt}
\begin{tabular}{lcccc}
\toprule
Method & Budget & Hit Rate & Cov. Rec. & Avg. Size \\
\midrule
Top-1 Fixed & 1 & 0.7727 & 0.7727 & 1.00 \\
\midrule
Top-3 Fixed & 3 & 0.8864 & \textbf{0.7008} & 3.00 \\
SkillSelect-Serve Compact@3 & 3 & \textbf{0.9091} & 0.6780 & 3.00 \\
\midrule
Top-5 Fixed & 5 & 0.9432 & \textbf{0.6936} & 5.00 \\
SkillSelect-Serve Final@5 & 5 & 0.9432 & 0.6621 & 5.00 \\
\midrule
Heuristic VarSize & adaptive & 0.9091 & 0.7583 & 3.10 \\
SkillSelect-Serve Aggressive@6 & 6 & \textbf{0.9659} & 0.6682 & 6.00 \\
\bottomrule
\end{tabular}
\end{table}

\begin{figure*}[t]
    \centering
    \includegraphics[width=0.92\textwidth]{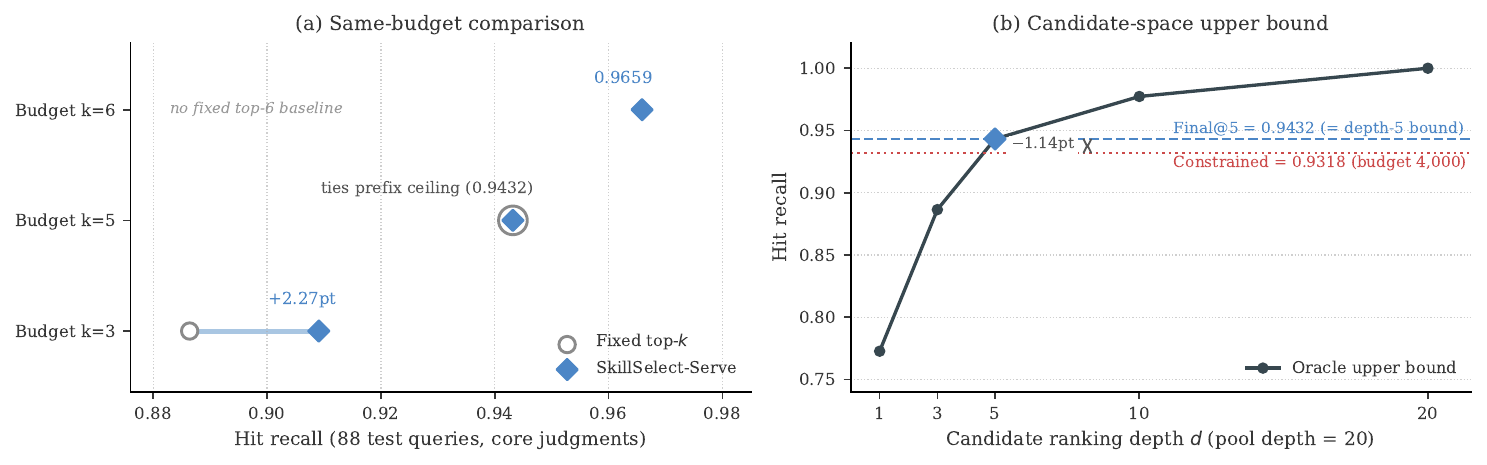}
    \caption{Result evidence board (88 held-out test queries, core multi-positive judgments). (a) Same-budget comparisons: hollow circles are fixed top-$k$ baselines and solid diamonds are SkillSelect-Serve operating points; Compact@3 improves hit rate by 2.27 points at an identical budget and Aggressive@6 reaches 0.9659. (b) Candidate-space upper bound versus ranking depth (judgments pooled to depth 20): Final@5 exactly attains the depth-5 bound of 0.9432 and the budget-constrained variant concedes only 1.14 points, indicating that residual error lies in discovery and ranking rather than in selection.}
    \label{fig:result_evidence_board}
\end{figure*}

Fig.~\ref{fig:result_evidence_board}(a) visualizes the same-budget comparisons in Table~\ref{tab:main_results}. In the compact setting with $k=3$, SkillSelect-Serve Compact@3 improves hit rate from 0.8864 (Top-3 Fixed) to 0.9091. Since the two methods use the same budget, this improvement cannot be attributed to loading more skills. Instead, it comes from marginal service-suitability estimation that surfaces critical services ranked below the retrieval prefix. Coverage recall of the prefix baseline is marginally higher (0.7008 versus 0.6780), which reflects a deliberate objective choice: the selector maximizes the probability of including at least one directly usable service---the event that determines whether the downstream agent can act at all---rather than accumulating interchangeable positives.

In the high-recall setting with $k=5$, Final@5 matches the prefix ceiling of the candidate ranking at 0.9432, and Aggressive@6 raises hit rate further to 0.9659. Heuristic VarSize attains a higher coverage recall largely because its smaller adaptive bundles shrink the normalization denominator; its hit rate remains below Aggressive@6 at a comparable trust level. SkillSelect-Serve is therefore best understood as a budget-controllable recommendation framework: Compact@3 for strict budgets, Final@5 as the default, and Aggressive@6 when a larger context cost is acceptable. Together with RQ1, this establishes the central quality claim of the paper: the QoS guarantees are obtained \emph{on top of} selection quality that matches or exceeds the fixed top-$k$ paradigm in hit rate at every service budget, not in exchange for it.

\smallskip
\noindent\fbox{\parbox{0.97\linewidth}{\textbf{Findings (RQ2):} At equal service budgets, budgeted selection matches or exceeds fixed top-$k$ retrieval in hit rate: it improves from 0.8864 to 0.9091 at $k{=}3$, matches the 0.9432 prefix ceiling at $k{=}5$, and reaches 0.9659 at $k{=}6$---so the QoS guarantees of RQ1 come on top of, not in exchange for, selection quality.}}

\subsection{RQ3: Which Modeling Granularity and Features Matter?}
\label{sec:ablation}

We answer RQ3 from two angles: which \emph{granularity} of utility modeling drives bundle quality, and which \emph{feature groups} of the service profile drive the marginal calibrator.
Table~\ref{tab:ablation_granularity} (top) presents the granularity ablation on test queries.

\begin{table}[t]
\centering
\caption{RQ3 ablations (88 test queries, core multi-positive judgments). Top: modeling granularity. Bottom: layered feature ablation of the marginal calibrator (ranking quality of individual-service relevance).}
\label{tab:ablation_granularity}
\footnotesize
\setlength{\tabcolsep}{4pt}
\begin{tabular}{lcccc}
\toprule
Method & Hit Rate & \multicolumn{2}{c}{Cov. Rec.} & Avg. Size \\
\midrule
Pure Neural Bundle Composer & 0.8295 & \multicolumn{2}{c}{0.5777} & 1.93 \\
Dual Bundle Calibrator & 0.8636 & \multicolumn{2}{c}{0.5549} & 2.69 \\
SkillSelect-Serve Compact@3 & 0.9091 & \multicolumn{2}{c}{0.6780} & 3.00 \\
SkillSelect-Serve Final@5 & 0.9432 & \multicolumn{2}{c}{0.6621} & 5.00 \\
SkillSelect-Serve Aggressive@6 & \textbf{0.9659} & \multicolumn{2}{c}{0.6682} & 6.00 \\
\midrule
Feature Layer & R@1 & R@5 & AP & $\Delta$AP \\
\midrule
L1 Retrieval only & 0.5455 & 0.7955 & 0.5458 & --- \\
L2 + Requirement match & 0.5795 & 0.8182 & 0.6177 & +0.0719 \\
L3 + Functional attrs & 0.5795 & 0.8295 & 0.6718 & +0.0541 \\
L4 + Cost/Risk (full) & \textbf{0.6023} & \textbf{0.8409} & \textbf{0.7069} & +0.0351 \\
\bottomrule
\end{tabular}
\end{table}

Pure Neural Bundle Composer achieves a hit rate of 0.8295, indicating that a neural bundle scorer alone is insufficient for reliably identifying critical services. Dual Bundle Calibrator improves hit rate to 0.8636 through bundle-level utility and hit calibration. The main improvement, however, comes from skill-level marginal service suitability: Compact@3 reaches 0.9091 with only three services, and Final@5 reaches 0.9432. SkillSelect-Serve thus adopts dual-granularity division of labor: skill-level marginal suitability (the deployed selection signal) identifies task-critical services, while bundle-level calibration complements it by estimating composition quality across coverage, redundancy, cost and risk.

\textbf{Which profile features carry the signal?} Table~\ref{tab:ablation_granularity} (bottom) ablates the marginal calibrator by incrementally enabling feature groups of the Skill Service Profile: (L1) retrieval signals only; (L2) adding requirement-matching features from the structured requirement object; (L3) adding functional attributes such as tool dependencies, input-output types, and tags; and (L4) adding cost and risk attributes.

The gains are monotonic and substantial: from L1 to L4, AP improves by 16.11 points (0.5458 to 0.7069) and Recall@1 by 5.68 points. The single largest contributor is the requirement-matching layer (+7.19 points of AP), which validates the design decision to interpose a structured requirement object between the raw query and the registry rather than matching on text similarity alone. Functional attributes and cost/risk attributes contribute a further +5.41 and +3.51 points, respectively---evidence that the service-profile abstraction, not merely stronger text ranking, is what the calibrator exploits.

Bundle-level hit rate is much less sensitive to these layers---at $k{=}5$ most configurations already reach the candidate-pool ceiling---so feature quality manifests primarily in \emph{where} relevant services sit in the ranking. This matters exactly when the budget is tight, and foreshadows the bottleneck analysis of RQ4.

\smallskip
\noindent\fbox{\parbox{0.97\linewidth}{\textbf{Findings (RQ3):} Calibrated skill-level marginal suitability drives selection quality (hit rate 0.9091 versus 0.8295--0.8636 for bundle-level scoring), and the service-profile abstraction carries the signal: requirement matching, functional attributes, and cost/risk features together add 16.11 points of AP over retrieval signals alone.}}

\subsection{RQ4: How Close Is SkillSelect-Serve to the Candidate-Space Upper Bound?}

To determine whether the residual error lies in candidate discovery or in bundle selection, we compute the candidate-space upper bound: the hit rate attainable if an oracle selected freely within the first $d$ candidates of the calibrated ranking. On the test queries the bound rises from 0.7727 at depth 1 to 0.8864 at depth 3, 0.9432 at depth 5, 0.9773 at depth 10, and 1.0000 at the pool depth of 20 (\textbf{full table is shown in supplementary material}).

Fig.~\ref{fig:result_evidence_board}(b) visualizes the source of the remaining gap. Final@5 exactly attains the depth-5 upper bound of 0.9432: within a five-service budget, selection extracts everything the top of the ranking offers, and even the budget-constrained variant concedes only 1.14 points against this bound. The queries that remain unrecovered at $k=5$ have their first relevant service at ranks 6--20; Aggressive@6 recovers a further 2.27 points by extending the budget one position. Improving the system beyond this point therefore requires promoting relevant services into the top-5 region---a discovery and ranking problem---rather than further tuning of the selection module.

Two qualifications apply. First, judgments are pooled to depth 20, so the depth-20 bound is exact by construction and is a within-pool bound rather than a registry-wide guarantee; the analysis cannot distinguish ranking depth beyond the pool. The bundle-level conclusions are unaffected, since all bundles select within the top six positions. Second, the upper bound is computed on the same calibrated ranking used by the selector, so it isolates the selection stage specifically: it shows that selection is near-optimal \emph{given} the ranking, and that the ranking itself is where the residual 5.7 points at $k=5$ reside. For a large-scale Skill Service Registry, system performance is jointly determined by discovery and selection; when the top-ranked region does not contain a relevant service, no downstream selection can recover it. Upper-bound analysis of this form provides a principled basis for locating modular bottlenecks in service computing systems.

\smallskip
\noindent\fbox{\parbox{0.97\linewidth}{\textbf{Findings (RQ4):} Selection is no longer the bottleneck: Final@5 exactly attains the depth-5 candidate-space bound of 0.9432, and the budget-constrained variant concedes only 1.14 points against it; further recall gains require promoting relevant services into the top-ranked region---a discovery and ranking problem.}}

\subsection{RQ5: What Happens in Downstream Small-Agent Execution?}

Finally, we conduct a diagnostic execution study relating offline recommendation quality to downstream small-agent execution utility; it is not intended as a state-of-the-art execution claim. The full per-method table is provided in the supplementary material.

Skill packages combined with the plan adapter improve execution: SkillSelect-Serve Compact@3 achieves a pass rate of 0.4000, above both the no-skill baseline (0.3143) and Top-3 retrieval (0.3714). However, Top-3 obtains a slightly higher average judge score, so better offline recommendation does not automatically translate into uniformly better execution. This recommendation--execution gap is expected: relevance-oriented labels measure whether a skill is related to a task, but not planning quality, prompt sensitivity, context interaction, or execution robustness. For service-oriented agent ecosystems, the implication is that service discovery and recommendation are necessary but not yet a complete execution loop; closing the gap requires execution-aware service utility learning, runtime feedback, and adaptive service reliability modeling. Representative success, counter, and context-pollution cases supporting this analysis are provided in the supplementary material.

\smallskip
\noindent\fbox{\parbox{0.97\linewidth}{\textbf{Findings (RQ5):} Recommended skill packages improve small-agent execution (pass rate 0.4000 versus 0.3143 without skills and 0.3714 for fixed Top-3 retrieval), but offline recommendation quality does not translate uniformly into execution utility; closing this gap requires execution-aware utility learning and runtime feedback.}}

\section{Discussion}
\label{sec:discussion}

\textbf{How should the QoS results be interpreted?} The three QoS dimensions behave differently and should be governed differently: the token budget is the load-bearing constraint that decides deliverability; risk is a soft, highly compressible dimension, because large registries contain low-risk substitutes for most capabilities, with hard exposure ceilings available when a deployment demands service-level guarantees; and tool availability is a hard environmental constraint whose price is unavoidable but explicit and monotone in environment scarcity. No fixed strategy is simultaneously optimal in recall, cost, risk, and redundancy. The role of the constrained projection is therefore not to output a universally optimal bundle, but to expose these dimensions as an operating interface that can be configured per deployment---strict sandbox, air-gapped, or full-privilege.

\textbf{What do the same-budget gains and the granularity ablation tell us?} The core benefit of SkillSelect-Serve comes from selecting better services under a fixed budget rather than from loading more skills (Tables~\ref{tab:selection_baselines} and~\ref{tab:main_results}). For small agents this matters operationally: every loaded service consumes context and may introduce conflicting procedures, so the appropriate evaluation is same-budget selection quality, not unconstrained recall. The granularity ablation locates the source of the gain: service identification and service composition are distinct decision problems, and the decisive factor is accurate identification of marginally useful services at the query-service level, with structured requirement matching contributing the single largest feature gain.

\textbf{Where is the bottleneck, and what does the execution diagnostic add?} Upper-bound analysis shows that the selection layer is no longer the bottleneck: Final@5 exactly attains the depth-5 bound of the calibrated ranking, so further recall improvement requires promoting relevant services into the top-ranked region---a discovery and ranking problem. The execution diagnostic adds a complementary boundary: offline relevance does not fully encode operability, composability, or contextual stability in an execution trajectory, so the recommendation layer, while necessary, is not yet a complete execution loop. Stronger discovery raises the candidate-space ceiling, and runtime feedback can turn delivered bundles into execution-aware QoS estimates---the two extensions that this modular decomposition makes precise.

\subsection{Limitations and Threats to Validity}

The conclusions should be interpreted within the following boundaries. First, the relevance judgments, although multi-positive and dual-assessor verified, are produced by LLM assessors over pools of depth 20. Inter-assessor agreement is substantial ($\kappa=0.62$--$0.70$) and the headline results are stable across assessors, but relevance grading remains subjective and services below the pooling depth are unjudged. Bundle-level conclusions are unaffected because all bundles select well within the pool, whereas the upper-bound curve is exact only up to the pooling depth. Similarly, the operational attribute labels come from a cost-efficient LLM annotator validated against a frontier reference (95.8\% adjacent agreement on risk); residual adjacent-level noise exists, which is why the risk mechanisms are evaluated on aggregated exposure rather than on individual label decisions.

Second, offline relevance is not equivalent to downstream execution utility, and the small-agent execution experiment is positioned strictly as a diagnostic rather than as an execution state-of-the-art claim. Third, performance is bounded by the candidate space and ranking of the discovery backbone: when a critical service does not reach the top-ranked region, no downstream selection can recover it under a tight budget, which is exactly what the upper-bound analysis quantifies. Fourth, the QoS results are operating trade-offs rather than a single point that dominates all metrics. Compact@3, Final@5, and Aggressive@6 correspond to different service-budget regimes. Finally, the main experiments follow a clean deployable protocol in which inference uses only label-free, deployment-observable features; oracle-derived utility appears solely in training supervision and upper-bound analysis, avoiding label leakage and keeping the results consistent with the actual deployment setting.

\section{Related Work}
\label{sec:related}

\textbf{Service computing, QoS-aware composition, and mashup recommendation.}
Service-oriented computing has long studied how distributed capabilities can be organized as discoverable, composable, and manageable services~\cite{papazoglou2007service,lemos2015web,bouguettaya2017service}. QoS-aware composition further establishes that selection must account for non-functional attributes---cost, latency, reliability, and user constraints---through global QoS utility~\cite{zeng2004qos}, collaborative QoS prediction~\cite{zheng2010qos}, and multi-objective composition optimization~\cite{zeng2025qos}, including resource- and context-constrained settings~\cite{njima2024web}. Mashup-oriented service recommendation is closest to our recommendation setting: service packages are selected from mashup descriptions~\cite{gu2016service}, CSBR models the compositional semantics of service bundles~\cite{gu2021csbr}, DySR captures service evolution with dynamic graph networks~\cite{liu2023dysr}, and diversity, compatibility, and graph structure are further exploited by a line of graph-based recommenders~\cite{gong2022dawar,wang2024sehgn,cao2024web,tang2024mashup,wang2025c,alhosaini2024api}. These lines provide the foundation for our Skill-as-a-Service formulation, but they target Web APIs and mashup co-invocation histories. Agent skills, by contrast, are consumed as \emph{context} by small LLM agents, so each selected service introduces token cost, tool dependencies, risk notes, and potential context interference.

\textbf{LLM-enhanced service recommendation.}
Recent work introduces LLMs into service recommendation for semantic representation and reasoning: LLM-enhanced representation learning for cold-start recommendation~\cite{rong2024llm}, service-network augmentation with GNN structural modeling~\cite{peng2025llmsrec}, LLM-augmented graph contrastive learning~\cite{zhu2025llm}, and MARS, which integrates semantic enrichment, structure-aware retrieval, and multi-agent reasoning over a constrained candidate space~\cite{liu2026mars}. These methods share our motivation that LLMs reduce the requirement--service semantic gap but require candidate constraints and modular decision-making for controllability; they remain Web-service-oriented, whereas our requirement planner only parses tasks into structured requirements and the final decision is made by deployable suitability estimation and constrained projection.

\textbf{LLM agents and tool/API use.}
Research on LLM tool use advances language models from text generation to environment interaction: ReAct interleaves reasoning with actions~\cite{yao2022react}, Toolformer learns self-supervised API calling~\cite{schick2023toolformer}, HuggingGPT orchestrates external model ecosystems~\cite{shen2023hugginggpt}, and ToolLLM/ToolBench and API-Bank provide large-scale tool-use data and evaluation~\cite{qin2024toolllm,li2023api}. Gorilla mitigates hallucination over massive APIs through retriever-aware training~\cite{patil2024gorilla}, RestGPT connects LLMs to RESTful APIs with coarse-to-fine planning~\cite{song2023restgpt}, iterative-feedback retrieval narrows the instruction--tool gap~\cite{xu2024enhancing}, and ToolGen unifies tool retrieval and calling as generation~\cite{wang2025toolgen}. These studies address \emph{how} LLMs use tools; this paper addresses the complementary question of \emph{which} Skill Services a small agent should load from a large library---a budgeted bundle decision rather than a single tool-call trajectory.

\textbf{Agent skill ecosystems and skill selection.}
A rapidly growing body of work treats reusable agent skills as first-class artifacts. SkillsBench evaluates curated skill packages on 86 tasks across 11 domains: curated skills raise average pass rates by 16.2 points but with high variance and negative deltas on 16 of 84 tasks~\cite{li2026skillsbench}---independently corroborating that the offline-relevance-versus-execution gap must be measured rather than assumed away. As registries grow, the selection problem itself has attracted dedicated methods: SkillRouter routes over an approximately 80K-skill registry with a compact retrieve-and-rerank pipeline~\cite{zheng2026skillrouter}; Graph-of-Skills performs budgeted selection over a typed dependency graph under a context budget~\cite{liu2026gos}; and the knapsack-based agent composer formulates component selection as constrained optimization via online sandbox probing~\cite{yuan2025knapsack}. These concurrent efforts confirm that large-scale skill selection is real and pressing, but none jointly enforces the three QoS dimensions studied here---a verified token budget, aggregated risk exposure, and deployment-environment tool availability---as first-class constraints within a single budgeted projection framework grounded in the service-computing tradition~\cite{zeng2004qos,zheng2010qos}. Specifically, SkillRouter stops at relevance ranking, Graph-of-Skills treats the context budget as its only operational constraint, the knapsack composer is not a deployable offline selector, and SkillsBench is a benchmark rather than a method.

\textbf{Retrieval, reranking, and bundle recommendation.}
BM25~\cite{robertson2009probabilistic}, dense sentence encoders~\cite{reimers2019sentence}, cross-encoder reranking~\cite{nogueira2019passage}, reciprocal rank fusion~\cite{cormack2009reciprocal}, and retrieval-augmented generation~\cite{lewis2020retrieval} are positioned in SkillSelect-Serve as the discovery stage: they supply a high-recall candidate pool but cannot decide whether services are redundant, complementary, too costly, or worth including from lower ranks. Bundle recommendation studies set-level selection~\cite{chang2020bundle,sun2024survey,he2022bundle}, and mashup API recommendation can be viewed as its service-domain instance~\cite{gu2016service,gu2021csbr,liu2023dysr}. We share the set-level perspective, and incorporate a skill bundle into a small agent's context that directly affects reasoning and execution paths, making the budget an operational deployment constraint instead of a ranking cutoff.

\textbf{Summary and positioning.}
As shown in Table~\ref{tab:related_work_comparison}, each related line covers part of the problem, but none combines a service-oriented skill model, budget control, first-class QoS constraints, and an explicitly measured recommendation--execution gap. SkillSelect-Serve unifies these concerns at the intersection of service computing and the LLM agent skill ecosystem, inheriting the discovery, composition, and QoS concerns of service-oriented computing while addressing the new constraints of skills consumed as agent context.

\begin{table*}[t]
\centering
\caption{Summary comparison of related work.}
\label{tab:related_work_comparison}
\small
\setlength{\tabcolsep}{3pt}
\renewcommand{\arraystretch}{1.15}
\begin{tabular}{p{0.50\textwidth}cccc}
\toprule
Approach Family
& Skill-as-Service
& Budget-control
& QoS-aware
& Rec.--Exec. Gap \\
\midrule
Service computing and QoS-aware composition~\cite{papazoglou2007service,lemos2015web,bouguettaya2017service,zeng2004qos,zeng2025qos,njima2024web}
& Partial & Partial & Yes & No \\
Mashup/API service recommendation~\cite{gu2016service,gu2021csbr,liu2023dysr,gong2022dawar,wang2024sehgn,cao2024web,tang2024mashup,wang2025c,alhosaini2024api}
& Partial & Limited & Partial & No \\
LLM-enhanced service recommendation~\cite{rong2024llm,peng2025llmsrec,zhu2025llm,liu2026mars}
& Partial & Limited & Partial & Limited \\
LLM tool/API use~\cite{yao2022react,schick2023toolformer,shen2023hugginggpt,qin2024toolllm,li2023api,patil2024gorilla,song2023restgpt,xu2024enhancing,wang2025toolgen}
& No & Limited & Limited & Execution-oriented \\
Agent skill routing and selection~\cite{li2026skillsbench,zheng2026skillrouter,liu2026gos,yuan2025knapsack}
& Partial & Partial & Limited & Partial \\
Retrieval/reranking methods~\cite{robertson2009probabilistic,reimers2019sentence,nogueira2019passage,cormack2009reciprocal,lewis2020retrieval}
& No & Top-$k$ only & No & No \\
Bundle recommendation~\cite{chang2020bundle,sun2024survey,he2022bundle}
& No & Partial & Limited & No \\
\midrule
\textbf{SkillSelect-Serve}
& Yes & Yes & Yes & Diagnostic \\
\bottomrule
\end{tabular}
\end{table*}

\section{Conclusion}
\label{sec:conclusion}

For small LLM agents, the key challenge of skill selection is how to select more useful, controllable, and deployable Skill Services under a limited operational budget, instead of retrieving more skills. This paper models agent skills as service-oriented capability units and formulates their selection as requirement-conditioned, budget-controllable, and QoS-aware Skill Service recommendation. SkillSelect-Serve realizes this formulation through requirement planning, candidate discovery, dual-granularity utility modeling, and budgeted QoS-aware projection---a deployable pipeline connecting the task-understanding capability of LLMs with the discovery, composition, and QoS-governance tradition of service computing.

Experiments on a 35,353-skill registry show that SkillSelect-Serve makes recommendations deliverable without sacrificing selection quality: enforcing the token budget converts feasibility from 9.1\% to 100\% at a cost of only 1.14 points of hit rate. Risk-aware selection halves delivered exposure, tool-availability gating eliminates the 44--81\% violation rates of tool-agnostic recommendation, and same-budget selection matches or exceeds fixed top-$k$ retrieval in hit rate at every service budget. The remaining boundaries are equally clear: further recall gains require stronger discovery, and a recommendation--execution gap remains between offline quality and downstream utility. Future work should learn execution-aware service utility from runtime feedback, closing the loop between recommendation and execution. This would turn skill libraries from passive document repositories into governable Skill Service Registries for large-scale reusable skill ecosystems.

\section*{Acknowledgements}
This research was supported by Natural Science Foundation of Zhejiang Province under Grant No.ZCLMS25F0201 and LZ25F020010,  Fundamental Research Funds for the Provincial Universities of Zhejiang under Grant No.GK259909299001-019, CCF-NetEase ThunderFire Innovation Research Funding under NO.CCF-Netease 202505, the National Natural Science Foundation of China under Grant No.62402151.

\bibliographystyle{IEEEtran}
\bibliography{References}

\begin{thebibliography}{10}
\providecommand{\url}[1]{#1}
\csname url@samestyle\endcsname
\providecommand{\newblock}{\relax}
\providecommand{\bibinfo}[2]{#2}
\providecommand{\BIBentrySTDinterwordspacing}{\spaceskip=0pt\relax}
\providecommand{\BIBentryALTinterwordstretchfactor}{4}
\providecommand{\BIBentryALTinterwordspacing}{\spaceskip=\fontdimen2\font plus
\BIBentryALTinterwordstretchfactor\fontdimen3\font minus
  \fontdimen4\font\relax}
\providecommand{\BIBforeignlanguage}[2]{{%
\expandafter\ifx\csname l@#1\endcsname\relax
\typeout{** WARNING: IEEEtran.bst: No hyphenation pattern has been}%
\typeout{** loaded for the language `#1'. Using the pattern for}%
\typeout{** the default language instead.}%
\else
\language=\csname l@#1\endcsname
\fi
#2}}
\providecommand{\BIBdecl}{\relax}
\BIBdecl

\bibitem{yao2022react}
S.~Yao, J.~Zhao, D.~Yu, N.~Du, I.~Shafran, K.~Narasimhan, and Y.~Cao, ``React:
  Synergizing reasoning and acting in language models,'' \emph{arXiv preprint
  arXiv:2210.03629}, 2022.

\bibitem{shen2023hugginggpt}
Y.~Shen, K.~Song, X.~Tan, D.~Li, W.~Lu, and Y.~Zhuang, ``Hugginggpt: Solving ai
  tasks with chatgpt and its friends in hugging face,'' \emph{Advances in
  Neural Information Processing Systems}, vol.~36, pp. 38\,154--38\,180, 2023.

\bibitem{schick2023toolformer}
T.~Schick, J.~Dwivedi-Yu, R.~Dess{\`\i}, R.~Raileanu, M.~Lomeli, E.~Hambro,
  L.~Zettlemoyer, N.~Cancedda, and T.~Scialom, ``Toolformer: Language models
  can teach themselves to use tools,'' \emph{Advances in neural information
  processing systems}, vol.~36, pp. 68\,539--68\,551, 2023.

\bibitem{qin2023toolllm}
Y.~Qin, S.~Liang, Y.~Ye, K.~Zhu, L.~Yan, Y.~Lu, Y.~Lin, X.~Cong, X.~Tang,
  B.~Qian \emph{et~al.}, ``Toolllm: Facilitating large language models to
  master 16000+ real-world apis,'' in \emph{The twelfth international
  conference on learning representations}, 2023.

\bibitem{li2023api}
M.~Li, Y.~Zhao, B.~Yu, F.~Song, H.~Li, H.~Yu, Z.~Li, F.~Huang, and Y.~Li,
  ``Api-bank: A comprehensive benchmark for tool-augmented llms,'' in
  \emph{Proceedings of the 2023 conference on empirical methods in natural
  language processing}, 2023, pp. 3102--3116.

\bibitem{papazoglou2007service}
M.~P. Papazoglou, P.~Traverso, S.~Dustdar, and F.~Leymann, ``Service-oriented
  computing: State of the art and research challenges,'' \emph{Computer},
  vol.~40, no.~11, pp. 38--45, 2007.

\bibitem{zeng2004qos}
L.~Zeng, B.~Benatallah, A.~H. Ngu, M.~Dumas, J.~Kalagnanam, and H.~Chang,
  ``Qos-aware middleware for web services composition,'' \emph{IEEE
  Transactions on software engineering}, vol.~30, no.~5, pp. 311--327, 2004.

\bibitem{bouguettaya2017service}
A.~Bouguettaya, M.~Singh, M.~Huhns, Q.~Z. Sheng, H.~Dong, Q.~Yu, A.~G. Neiat,
  S.~Mistry, B.~Benatallah, B.~Medjahed \emph{et~al.}, ``A service computing
  manifesto: the next 10 years,'' \emph{Communications of the ACM}, vol.~60,
  no.~4, pp. 64--72, 2017.

\bibitem{zheng2010qos}
Z.~Zheng, H.~Ma, M.~R. Lyu, and I.~King, ``Qos-aware web service recommendation
  by collaborative filtering,'' \emph{IEEE Transactions on Services Computing},
  vol.~4, no.~2, pp. 140--152, 2011.

\bibitem{voorhees2005trec}
E.~M. Voorhees and D.~K. Harman, \emph{TREC: Experiment and evaluation in
  information retrieval}.\hskip 1em plus 0.5em minus 0.4em\relax MIT Press,
  2005.

\bibitem{carbonell1998mmr}
J.~Carbonell and J.~Goldstein, ``The use of mmr, diversity-based reranking for
  reordering documents and producing summaries,'' in \emph{Proceedings of the
  21st annual international ACM SIGIR conference on Research and development in
  information retrieval}, 1998, pp. 335--336.

\bibitem{robertson2009probabilistic}
S.~Robertson and H.~Zaragoza, \emph{The probabilistic relevance framework: BM25
  and beyond}.\hskip 1em plus 0.5em minus 0.4em\relax Now Publishers Inc, 2009,
  vol.~4.

\bibitem{reimers2019sentence}
N.~Reimers and I.~Gurevych, ``Sentence-bert: Sentence embeddings using siamese
  bert-networks,'' in \emph{Proceedings of the 2019 conference on empirical
  methods in natural language processing and the 9th international joint
  conference on natural language processing (EMNLP-IJCNLP)}, 2019, pp.
  3982--3992.

\bibitem{cormack2009reciprocal}
G.~V. Cormack, C.~L. Clarke, and S.~Buettcher, ``Reciprocal rank fusion
  outperforms condorcet and individual rank learning methods,'' in
  \emph{Proceedings of the 32nd international ACM SIGIR conference on Research
  and development in information retrieval}, 2009, pp. 758--759.

\bibitem{nogueira2019passage}
R.~Nogueira and K.~Cho, ``Passage re-ranking with bert,'' \emph{arXiv preprint
  arXiv:1901.04085}, 2019.

\bibitem{friedman2001greedy}
J.~H. Friedman, ``Greedy function approximation: a gradient boosting machine,''
  \emph{Annals of statistics}, pp. 1189--1232, 2001.

\bibitem{lemos2015web}
A.~L. Lemos, F.~Daniel, and B.~Benatallah, ``Web service composition: a survey
  of techniques and tools,'' \emph{ACM Computing Surveys (CSUR)}, vol.~48,
  no.~3, pp. 1--41, 2015.

\bibitem{zeng2025qos}
L.~Zeng, B.~Benatallah, M.~Dumas, J.~Kalagnanam, and A.~H. Ngu, ``Qos-aware
  service composition: A retrospective,'' \emph{IEEE Transactions on Software
  Engineering}, vol.~51, no.~3, pp. 836--841, 2025.

\bibitem{njima2024web}
C.~B. Njima, C.~G. Guegan, Y.~Gamha, and L.~B. Romdhane, ``Web service
  composition in mobile environment: a survey of techniques,'' \emph{IEEE
  Transactions on Services Computing}, vol.~17, no.~2, pp. 689--704, 2024.

\bibitem{gu2016service}
Q.~Gu, J.~Cao, and Q.~Peng, ``Service package recommendation for mashup
  creation via mashup textual description mining,'' in \emph{2016 IEEE
  international conference on web services (ICWS)}.\hskip 1em plus 0.5em minus
  0.4em\relax IEEE, 2016, pp. 452--459.

\bibitem{gu2021csbr}
Q.~Gu, J.~Cao, and Y.~Liu, ``Csbr: A compositional semantics-based service
  bundle recommendation approach for mashup development,'' \emph{IEEE
  Transactions on Services Computing}, vol.~15, no.~6, pp. 3170--3183, 2021.

\bibitem{liu2023dysr}
M.~Liu, Z.~Tu, H.~Xu, X.~Xu, and Z.~Wang, ``Dysr: A dynamic graph neural
  network based service bundle recommendation model for mashup creation,''
  \emph{IEEE Transactions on Services Computing}, vol.~16, no.~4, pp.
  2592--2605, 2023.

\bibitem{gong2022dawar}
W.~Gong, X.~Zhang, Y.~Chen, Q.~He, A.~Beheshti, X.~Xu, C.~Yan, and L.~Qi,
  ``Dawar: Diversity-aware web apis recommendation for mashup creation based on
  correlation graph,'' in \emph{Proceedings of the 45th international ACM SIGIR
  conference on Research and Development in information retrieval}, 2022, pp.
  395--404.

\bibitem{wang2024sehgn}
X.~Wang, M.~Xi, Y.~Li, X.~Pan, Y.~Wu, S.~Deng, and J.~Yin, ``Sehgn:
  Semantic-enhanced heterogeneous graph network for web api recommendation,''
  \emph{IEEE Transactions on Services Computing}, vol.~17, no.~5, pp.
  2836--2849, 2024.

\bibitem{cao2024web}
B.~Cao, Q.~Peng, X.~Xie, Z.~Peng, J.~Liu, and Z.~Zheng, ``Web service
  recommendation via combining topic-aware heterogeneous graph representation
  and interactive semantic enhancement,'' \emph{IEEE Transactions on Services
  Computing}, vol.~17, no.~6, pp. 4451--4466, 2024.

\bibitem{tang2024mashup}
M.~Tang, F.~Xie, S.~Lian, J.~Mai, and S.~Li, ``Mashup-oriented api
  recommendation via pre-trained heterogeneous information networks,''
  \emph{Information and Software Technology}, vol. 169, p. 107428, 2024.

\bibitem{wang2025c}
Y.~Wang, L.~Yang, W.~Gong, M.~Khosravi, M.~Khan, and W.~Rafique, ``C-da w ar:
  Towards diversity-aware web apis recommendation for mashup creation based on
  contrastive learning,'' \emph{Tsinghua Science and Technology}, 2025.

\bibitem{alhosaini2024api}
H.~Alhosaini, S.~Alharbi, X.~Wang, and G.~Xu, ``Api recommendation for mashup
  creation: A comprehensive survey,'' \emph{The Computer Journal}, vol.~67,
  no.~5, pp. 1920--1940, 2024.

\bibitem{rong2024llm}
D.~Rong, L.~Yao, Y.~Zheng, S.~Yu, X.~Xu, M.~Liu, and Z.~Wang, ``Llm enhanced
  representation for cold start service recommendation,'' in
  \emph{International Conference on Service-Oriented Computing}.\hskip 1em plus
  0.5em minus 0.4em\relax Springer, 2024, pp. 153--167.

\bibitem{peng2025llmsrec}
Q.~Peng, B.~Cao, X.~Xie, H.~Ye, J.~Liu, and Z.~Li, ``Llmsrec: Large language
  model with service network augmentation for web service recommendation,''
  \emph{Knowledge-Based Systems}, vol. 323, p. 113710, 2025.

\bibitem{zhu2025llm}
Y.~Zhu, Z.~Lin, J.~Fan, M.~Liu, and Z.~Wang, ``Llm-cosr: Noise-resistant
  service recommendation via llm-augmented graph contrastive learning,'' in
  \emph{2025 IEEE International Conference on Web Services (ICWS)}.\hskip 1em
  plus 0.5em minus 0.4em\relax IEEE, 2025, pp. 1--10.

\bibitem{liu2026mars}
M.~Liu, Z.~Yin, C.~Tian, S.~Yu, T.~Cai, Z.~Xu, and Z.~Wang, ``Mars: A
  multi-agent collaborative reasoning framework for service recommendation,''
  \emph{IEEE Transactions on Services Computing}, 2026.

\bibitem{qin2024toolllm}
Y.~Qin, S.~Liang, Y.~Ye, K.~Zhu, L.~Yan, Y.~Lu, Y.~Lin, X.~Cong, X.~Tang,
  B.~Qian \emph{et~al.}, ``Toolllm: Facilitating large language models to
  master 16000+ real-world apis,'' in \emph{International Conference on
  Learning Representations}, vol. 2024, 2024, pp. 9695--9717.

\bibitem{patil2024gorilla}
S.~G. Patil, T.~Zhang, X.~Wang, and J.~E. Gonzalez, ``Gorilla: Large language
  model connected with massive apis,'' \emph{Advances in Neural Information
  Processing Systems}, vol.~37, pp. 126\,544--126\,565, 2024.

\bibitem{song2023restgpt}
Y.~Song, W.~Xiong, D.~Zhu, W.~Wu, H.~Qian, M.~Song, H.~Huang, C.~Li, K.~Wang,
  R.~Yao \emph{et~al.}, ``Restgpt: Connecting large language models with
  real-world restful apis,'' \emph{arXiv preprint arXiv:2306.06624}, 2023.

\bibitem{xu2024enhancing}
Q.~Xu, Y.~Li, H.~Xia, and W.~Li, ``Enhancing tool retrieval with iterative
  feedback from large language models,'' in \emph{Findings of the Association
  for Computational Linguistics: EMNLP 2024}, 2024, pp. 9609--9619.

\bibitem{wang2025toolgen}
R.~Wang, X.~Han, L.~Ji, S.~Wang, T.~Baldwin, and H.~Li, ``Toolgen: Unified tool
  retrieval and calling via generation,'' in \emph{International Conference on
  Learning Representations}, vol. 2025, 2025, pp. 73\,473--73\,498.

\bibitem{li2026skillsbench}
X.~Li, W.~Chen, Y.~Liu, S.~Zheng, X.~Chen, Y.~He \emph{et~al.}, ``Skillsbench:
  Benchmarking how well agent skills work across diverse tasks,'' \emph{arXiv
  preprint arXiv:2602.12670}, 2026.

\bibitem{zheng2026skillrouter}
Y.~Zheng, Z.~Zhang, C.~Ma, Y.~Yu, J.~Zhu, W.~Yong, T.~Xu, B.~Dong, H.~Zhu,
  R.~Huang, and G.~Yu, ``Skillrouter: Skill routing for llm agents at scale,''
  \emph{arXiv preprint arXiv:2603.22455}, 2026.

\bibitem{liu2026gos}
D.~Liu, Z.~Li, H.~Du, X.~Wu, S.~Gui, Y.~Kuang, and L.~Sun, ``Graph-of-skills:
  Dependency-aware structural retrieval for massive agent skills,'' \emph{arXiv
  preprint arXiv:2604.05333}, 2026.

\bibitem{yuan2025knapsack}
M.~Yuan, K.~Pahwa, S.~Chang, M.~D. Kaba, J.~Jiang, X.~Ma, Y.~Zhang, and
  M.~Sunkara, ``Automated composition of agents: A knapsack approach for
  agentic component selection,'' in \emph{Advances in Neural Information
  Processing Systems}, vol.~38, 2025.

\bibitem{lewis2020retrieval}
P.~Lewis, E.~Perez, A.~Piktus, F.~Petroni, V.~Karpukhin, N.~Goyal,
  H.~K{\"u}ttler, M.~Lewis, W.-t. Yih, T.~Rockt{\"a}schel \emph{et~al.},
  ``Retrieval-augmented generation for knowledge-intensive nlp tasks,''
  \emph{Advances in neural information processing systems}, vol.~33, pp.
  9459--9474, 2020.

\bibitem{chang2020bundle}
J.~Chang, C.~Gao, X.~He, D.~Jin, and Y.~Li, ``Bundle recommendation with graph
  convolutional networks,'' in \emph{Proceedings of the 43rd international ACM
  SIGIR conference on Research and development in Information Retrieval}, 2020,
  pp. 1673--1676.

\bibitem{sun2024survey}
M.~Sun, L.~Li, M.~Li, X.~Tao, D.~Zhang, Q.~Xie, P.~Wang, and J.~X. Huang, ``A
  survey on bundle recommendation: Methods, applications, and challenges,''
  \emph{ACM Computing Surveys}, 2024.

\bibitem{he2022bundle}
Z.~He, H.~Zhao, T.~Yu, S.~Kim, F.~Du, and J.~McAuley, ``Bundle mcr: Towards
  conversational bundle recommendation,'' in \emph{Proceedings of the 16th ACM
  Conference on Recommender Systems}, 2022, pp. 288--298.

\end{thebibliography}

\clearpage
\appendices

\section{Correspondence with Service-Oriented Architecture Concepts}
\label{supp:soa}

This section makes explicit how the components of SkillSelect-Serve instantiate classical service-oriented architecture (SOA) roles. Table~\ref{tab:soa_mapping} summarizes the correspondence.

\begin{table}[h]
\centering
\caption{Mapping between classical SOA concepts and their instantiation in SkillSelect-Serve.}
\label{tab:soa_mapping}
\footnotesize
\setlength{\tabcolsep}{3pt}
\renewcommand{\arraystretch}{1.15}
\begin{tabular}{p{0.34\linewidth}p{0.58\linewidth}}
\toprule
SOA Concept & Instantiation in This Work \\
\midrule
Service provider & Skill author / upstream skill repository \\
Service consumer & Small LLM agent operating under context, tool, and risk constraints \\
Service registry & Skill Service Registry of 35,353 profiled skills \\
Service description & Skill Service Profile: capability summary, input-output assumptions, required tools, context cost, risk level \\
Requirement specification & Structured service requirement object produced by the requirement planner \\
Service discovery & Requirement-conditioned candidate retrieval (lexical, dense, rank fusion, reranking) \\
Service selection & Calibrated task-conditioned suitability ranking \\
Service composition & Budgeted bundle packing over the suitability ranking \\
QoS constraints & Token budget, aggregated risk budget, tool availability, task-level hard constraints \\
QoS governance & Risk gating and tool-availability gating with per-query feasibility guarantees \\
\bottomrule
\end{tabular}
\end{table}

The correspondence is deliberately scoped to what the framework currently implements. Classical SOA capabilities that are \emph{not} yet covered---explicit orchestration ordering among selected services, interface adapters for input-output mismatches, and runtime SLA monitoring with failure recovery---are natural extensions of the Skill Service Profile and are left to future work, consistent with the execution-aware direction discussed in the main paper.

\section{Operational Attribute Annotation Validation}
\label{supp:annotation}

The two operational attributes of the Skill Service Profile that require semantic judgment---the risk level and the required-tool set---are annotated by an LLM over the full registry of 35,353 skills. This section documents the validation protocol summarized in the main paper, followed by a blind two-annotator human study covering both the risk labels and the relevance judgments used as evaluation gold.

\subsection{Validation Against a Frontier Reference}

We deliberately avoid keyword heuristics for these attributes. On a stratified 120-skill sample judged by a frontier model, keyword-based risk extraction agrees with the reference on only 34.2\% of skills and systematically over-reports high risk, because innocuous terms such as ``token'' trigger security patterns. The production annotator is a cost-efficient LLM validated against the frontier reference on the same held-out sample, reaching 95.8\% adjacent agreement (71.7\% exact) on the four-level risk scale. Disagreements are concentrated in adjacent levels (e.g., low versus medium) rather than in the high-risk decisions that drive gating. The resulting labels exhibit a plausible long-tail risk distribution, with only 1.7\% of skills labeled high-risk.

\subsection{Blind Two-Annotator Human Study}
\label{supp:human_study}

To measure how well the LLM annotations track human judgment, two graduate annotators independently labeled two blind sheets, without access to each other, to any model output, or to the answer keys: (i) 300 query--skill relevance pairs (30 test queries $\times$ 10 candidates, candidate order shuffled) on the same three-level scale used by the LLM judges (2 = core, 1 = helpful, 0 = irrelevant); and (ii) 120 stratified skills on the four-level risk scale. Sampling is seeded and reproducible; the sheets, guideline, and analysis script are included in the public repository. Table~\ref{tab:human_study} reports exact agreement, adjacent ($\pm$1 level) agreement, Cohen's $\kappa$, and linearly weighted $\kappa$; the human consensus is the rounded mean of the two annotators.

\begin{table}[h]
\centering
\caption{Blind human study: inter-annotator ceiling and human--LLM agreement on relevance grades ($n{=}300$) and risk levels ($n{=}120$).}
\label{tab:human_study}
\footnotesize
\setlength{\tabcolsep}{3pt}
\renewcommand{\arraystretch}{1.15}
\begin{tabular}{llcccc}
\toprule
Task & Comparison & Exact & Adj.\ ($\pm$1) & $\kappa$ & $\kappa_{\mathrm{lin}}$ \\
\midrule
Relevance & Human A vs.\ Human B & 0.977 & 1.000 & 0.965 & 0.973 \\
 & Consensus vs.\ primary judge & 0.640 & 0.993 & 0.462 & 0.586 \\
 & Consensus vs.\ second assessor & 0.670 & 0.967 & 0.510 & 0.604 \\
\midrule
Risk & Human A vs.\ Human B & 0.950 & 1.000 & 0.927 & 0.948 \\
 & Consensus vs.\ production annotator & 0.708 & 0.933 & 0.572 & 0.632 \\
 & Consensus vs.\ frontier reference & 0.667 & 0.917 & 0.512 & 0.593 \\
\bottomrule
\end{tabular}
\end{table}

Three observations follow. \emph{First}, inter-annotator agreement is almost perfect ($\kappa = 0.965$ and $0.927$; adjacent agreement 100\%), confirming that both annotation tasks are well defined and establishing a human ceiling. \emph{Second}, human--LLM agreement is moderate to substantial at the exact level but near-ceiling at the adjacent level (93--99\%): disagreements are almost exclusively one-level boundary cases, never severe conflations of irrelevant with core or benign with high-risk. \emph{Third}, the residual disagreement is systematic rather than random, and its direction is conservative for our conclusions. On relevance, 100 of the 108 consensus-level disagreements with the primary judge are cases where humans grade \emph{higher} than the LLM; on the binary positive labels that the evaluation actually consumes, humans confirm nearly all LLM positives (recall 0.92 for core and 0.97 for relaxed positives; $\kappa = 0.62$ and $0.60$) while additionally accepting some candidates the LLM rejects. The LLM-derived gold is thus a strict subset of what humans consider relevant, so the reported hit-rate and coverage numbers are lower bounds and are not inflated by the automatic judging. On risk, all disagreements stay within one level, and the rare human high-risk calls (6 of 120) fall on skills the production annotator rates medium, i.e., adjacent to the gating boundary rather than mislabeled as benign.

\section{Full QoS Constraint Tables (RQ1)}
\label{supp:qos}

This section provides the full per-setting tables behind the risk and tool-availability results summarized in RQ1 of the main paper; the corresponding operating frontiers are visualized in Fig.~2 of the main text.

Table~\ref{tab:qos_risk} reports the complete recall--risk trade-off. Risk-aware ranking subtracts a risk penalty $\lambda$ from the calibrated score before selection; the aggregated risk budget enforces a per-bundle exposure ceiling inside the constrained projection.

\begin{table}[h]
\centering
\caption{Recall--risk trade-off (test queries). Top: risk-aware ranking at $k{=}5$. Bottom: aggregated risk budget inside the constrained projection; ``would violate'' is the fraction of queries whose risk-agnostic bundle exceeds the ceiling.}
\label{tab:qos_risk}
\footnotesize
\setlength{\tabcolsep}{3pt}
\begin{tabular}{lcccc}
\toprule
Setting & Hit Rate & Exposure & $\Delta$Exposure & Would violate \\
\midrule
$\lambda=0$ (risk-agnostic) & 0.9432 & 1.475 & --- & --- \\
$\lambda=0.05$ & 0.9318 & 0.694 & $-53.0\%$ & --- \\
$\lambda=0.10$ & 0.9318 & 0.588 & $-60.1\%$ & --- \\
$\lambda=0.15$ & 0.9205 & 0.530 & $-64.1\%$ & --- \\
\midrule
No risk budget & 0.9318 & 1.390 & --- & --- \\
Risk budget 1.50 & 0.9318 & 1.077 & $-22.5\%$ & 48.9\% \\
Risk budget 1.25 & 0.9318 & 0.907 & $-34.7\%$ & 57.9\% \\
Risk budget 1.00 & 0.8977 & 0.725 & $-47.8\%$ & 76.1\% \\
Risk budget 0.75 & 0.8977 & 0.584 & $-58.0\%$ & 86.4\% \\
\bottomrule
\end{tabular}
\end{table}

Table~\ref{tab:qos_tools} reports tool-availability gating across five deployment environments defined over the 12-tool vocabulary.

\begin{table*}[t]
\centering
\caption{Tool-availability gating across deployment environments (test queries, constrained projection, $k{=}5$). Violation is the fraction of bundles containing a service whose required tools are unavailable; footprint is the average number of tool-dependent services per bundle.}
\label{tab:qos_tools}
\begin{tabular}{lccccc}
\toprule
Environment & Available Tools & Violation (tool-agnostic) & Violation (gated) & Hit Rate (gated) & Tool Footprint \\
\midrule
Full toolset & 12 & 0.0\% & 0.0\% & 0.9318 & 2.55 \\
No network & 11 & 44.3\% & \textbf{0.0\%} & 0.8750 & 1.91 \\
No shell & 11 & 58.0\% & \textbf{0.0\%} & 0.7955 & 1.64 \\
No execution, no network & 10 & 73.9\% & \textbf{0.0\%} & 0.7500 & 1.11 \\
Python sandbox & 3 & 80.7\% & \textbf{0.0\%} & 0.7273 & 0.74 \\
\bottomrule
\end{tabular}
\end{table*}

\section{Candidate-Space Upper-Bound Table (RQ4)}
\label{supp:upper_bound}

Table~\ref{tab:upper_bound} reports the full candidate-space upper-bound analysis referenced in RQ4 of the main paper: the hit rate attainable if an oracle selected freely within the first $d$ candidates of the calibrated ranking, alongside the operating points of SkillSelect-Serve.

\begin{table}[h]
\centering
\caption{Candidate-space upper-bound analysis (88 test queries). Relevance judgments are pooled to depth 20, so the upper bound at depth 20 is exact by construction and depths beyond 20 cannot register additional unjudged positives.}
\label{tab:upper_bound}
\begin{tabular}{lc}
\toprule
Candidate Space / Method & Hit Rate \\
\midrule
Top-1 candidate & 0.7727 \\
Top-3 candidates & 0.8864 \\
Top-5 candidates & 0.9432 \\
Top-10 candidates & 0.9773 \\
Top-20 candidates (pool depth) & 1.0000 \\
\midrule
SkillSelect-Serve (constrained, budget 4{,}000) & 0.9318 \\
SkillSelect-Serve Final@5 & 0.9432 \\
SkillSelect-Serve Aggressive@6 & 0.9659 \\
\bottomrule
\end{tabular}
\end{table}

\section{Statistical Significance of Headline Comparisons}
\label{supp:significance}

The main paper reports exact metric values because the deployed pipeline is deterministic at inference time. This section quantifies the sampling uncertainty of the headline hit-rate comparisons over the 88 test queries. For each pair of methods we report the exact two-sided McNemar test on discordant queries and a paired bootstrap (10{,}000 resamples) of the hit-rate difference; per-method bootstrap 95\% confidence intervals are also given. Table~\ref{tab:significance} summarizes the results.

\begin{table}[h]
\centering
\caption{Paired significance analysis of hit-rate comparisons (88 test queries, core multi-positive judgments). Discordant counts are (A hits, B misses) / (A misses, B hits).}
\label{tab:significance}
\footnotesize
\setlength{\tabcolsep}{2pt}
\renewcommand{\arraystretch}{1.12}
\begin{tabular}{p{0.31\linewidth}cccc}
\toprule
Comparison (A vs.\ B) & $\Delta$ & Disc. & McNemar $p$ & 95\% CI \\
\midrule
Compact@3 vs.\ Top-3 Fixed & $+0.0227$ & 2/0 & 0.50 & $[0.000, +0.057]$ \\
Final@5 vs.\ Top-5 Fixed & $0.0000$ & 1/1 & 1.00 & $[-0.034, +0.034]$ \\
Aggressive@6 vs.\ Top-5 Fixed & $+0.0227$ & 2/0 & 0.50 & $[0.000, +0.057]$ \\
Constrained@5 vs.\ Final@5 & $-0.0114$ & 0/1 & 1.00 & $[-0.034, 0.000]$ \\
\bottomrule
\end{tabular}
\end{table}

Three observations follow. First, the same-budget hit-rate differences correspond to only one or two discordant queries and are not statistically significant at this test-set size; the direction is nevertheless consistent (SkillSelect-Serve never loses a query that the fixed-$k$ baseline recovers, except one at $k{=}5$ where the reverse also occurs once). Second, and symmetrically, the 1.14-point constraint cost of the budgeted projection is statistically indistinguishable from zero ($p{=}1.00$, CI $[-0.034, 0.000]$), supporting the claim that deliverability is obtained at minimal relevance loss. Third, the load-bearing QoS results of RQ1 are of a different nature: 100\% token-budget feasibility and 0\% tool-violation rates hold \emph{by construction} on every query, so they are guarantees of the constrained projection rather than sampled estimates subject to hypothesis testing. Per-method hit rates with bootstrap 95\% CIs: Top-3 Fixed 0.8864 $[0.818, 0.943]$; Compact@3 0.9091 $[0.841, 0.966]$; Top-5 Fixed and Final@5 0.9432 $[0.886, 0.989]$; Constrained@5 0.9318 $[0.875, 0.977]$; Aggressive@6 0.9659 $[0.920, 1.000]$.

\section{Exact and Beam-Search Selection Baselines}
\label{supp:exact_selector}

The main paper describes the constrained projection as a greedy procedure that guarantees feasibility by construction without claiming global optimality. This section quantifies the gap to the optimum. We solve the deployed selection problem \emph{exactly} with a branch-and-bound solver (equivalent to an integer linear program over the candidate pool: maximize the sum of calibrated marginal suitability subject to the 4{,}000-token budget, at most one high-risk service, and $k\leq 5$), and additionally run a beam-search selector (width 50) over the same feasible set. All three selectors consume identical candidate pools, scores, and constraint definitions; the greedy selector is re-derived in-script and verified to reproduce the published constrained bundles exactly on all 577 queries.

\begin{table}[h]
\centering
\caption{Greedy projection versus exact and beam-search selection under the deployed feasible set (test queries; selection time averaged over all 577 queries).}
\label{tab:exact_selector}
\footnotesize
\setlength{\tabcolsep}{2.5pt}
\begin{tabular}{lcccc}
\toprule
Selector & Hit Rate & Cov.\ Rec. & Feasible & ms/query \\
\midrule
Greedy projection (deployed) & 0.9318 & 0.6053 & 100\% & 0.005 \\
Exact optimum (B\&B / ILP) & 0.9318 & 0.6053 & 100\% & 0.222 \\
Beam search (width 50) & 0.9318 & 0.6053 & 100\% & 0.696 \\
\bottomrule
\end{tabular}
\end{table}

The greedy projection attains the exact optimum of the stated objective on 89.1\% of all 577 queries, and on the 88 test queries the exact and beam-search selectors produce \emph{identical} hit rate and coverage recall to the greedy projection (coverage is recomputed here under a single shared protocol for the three selectors). The residual objective gap on the remaining queries therefore never changes a recall outcome, while the greedy selector runs roughly $44\times$ faster than the exact solver. This certifies empirically that the skip-and-continue projection, though not provably optimal, leaves no measurable recall on the table under the deployed constraint configuration.

\section{Diagnostic Small-Agent Execution Study (RQ5)}
\label{supp:execution}

Table~\ref{tab:diagnostic_execution} reports the full results of the diagnostic execution study referenced in the main paper. The study relates offline recommendation quality to downstream small-agent execution utility and is not intended as a state-of-the-art execution claim.

\begin{table}[h]
\centering
\caption{Diagnostic small-agent execution study.}
\label{tab:diagnostic_execution}
\small
\setlength{\tabcolsep}{3pt}
\renewcommand{\arraystretch}{1.12}
\begin{tabular}{p{0.36\linewidth}ccp{0.26\linewidth}}
\toprule
Method & Pass Rate & Score & Observation \\
\midrule
No Skill
& 0.3143 & --
& Lower bound \\
Top-3 + Plan Adapter
& 0.3714 & 1.9143
& Strong retrieval baseline \\
SkillSelect-Serve Compact@3 + Plan Adapter
& 0.4000 & 1.8571
& Higher pass rate \\
\bottomrule
\end{tabular}
\end{table}

Skill packages combined with the plan adapter improve execution: SkillSelect-Serve Compact@3 achieves a pass rate of 0.4000, above both the no-skill baseline (0.3143) and Top-3 retrieval (0.3714). However, Top-3 obtains a slightly higher average judge score, so better offline recommendation does not automatically translate into uniformly better execution.

\section{Case Studies}
\label{supp:cases}

To further understand the execution-study observations, we analyze several representative cases.

\textbf{Success case.} In this task, fixed Top-3 retrieval returns multiple skill services that are semantically similar but functionally redundant. In contrast, SkillSelect-Serve Compact@3 selects a lower-ranked service with critical marginal usefulness, thereby supplying the file-processing or data-transformation capability required by the task. This case shows that marginal service suitability can identify key services that are easily overlooked by retrieval rank alone.

\textbf{Counter-case.} In some execution tasks, Top-3 retrieval returns a generic executable service that may not achieve the highest offline utility, but can still provide a more direct execution path for the small agent when combined with the plan adapter. This explains why Top-3 + Plan Adapter remains competitive in terms of judge score.

\textbf{Context pollution case.} In high-recall settings, Final@5 or Aggressive@6 can improve bundle recall, but the extra services they introduce may increase the context burden and interfere with small-agent execution. This further indicates that budgeted recommendation should incorporate execution-aware feedback, rather than optimizing offline recall alone.

Overall, the case studies are consistent with the quantitative results. SkillSelect-Serve better identifies critical Skill Services under the same budget and approaches the candidate-space upper bound in the high-recall setting, while downstream execution is additionally affected by planning, adaptation, and context interaction. These findings motivate future research on execution-aware Skill Service Composition.

\section{Additional Architecture Diagrams}
\label{supp:figures}

\begin{figure}[p]
    \centering
    \includegraphics[width=\linewidth]{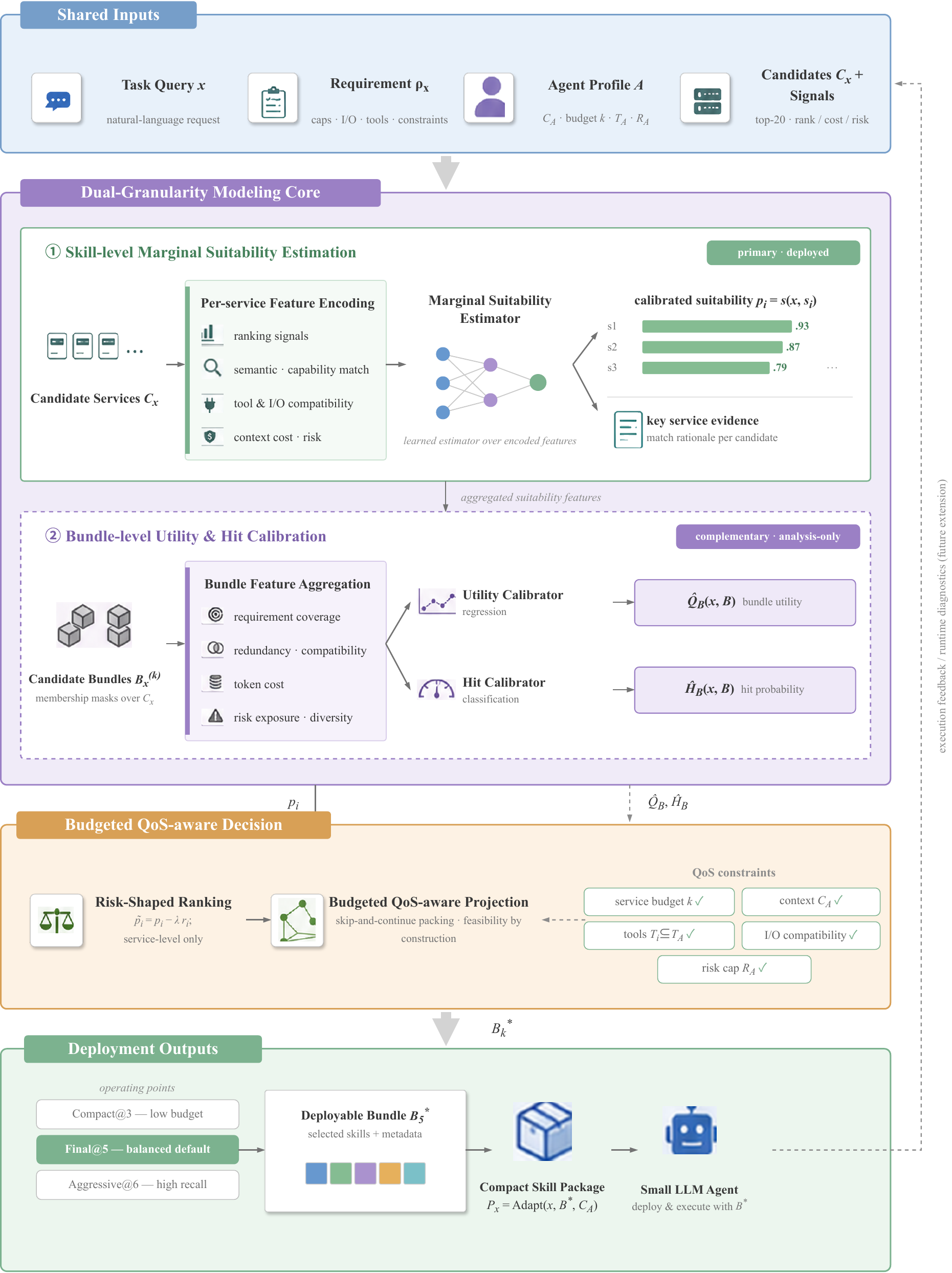}
    \caption{Service-aware dual-granularity modeling: the service-level suitability branch (primary, deployed) and the bundle-level quality/hit calibration branch (complementary, analysis-oriented). This figure expands Stage~3 (dual-granularity scoring) of the overview figure (Fig.~1) in the main text.}
    \label{fig:dual_granularity_utility}
\end{figure}

Fig.~\ref{fig:dual_granularity_utility} details the service-aware dual-granularity modeling module. The service-level branch---task-conditioned service suitability estimation---is the primary decision signal whose calibrated scores directly drive the deployed constrained selection. The bundle-level branch---quality and hit calibration---estimates requirement coverage, compatibility, complementarity, redundancy, context cost, and risk exposure of a candidate bundle, and supports composition-quality analysis and the granularity-ablation variants.

\begin{figure}[p]
    \centering
    \includegraphics[width=\linewidth]{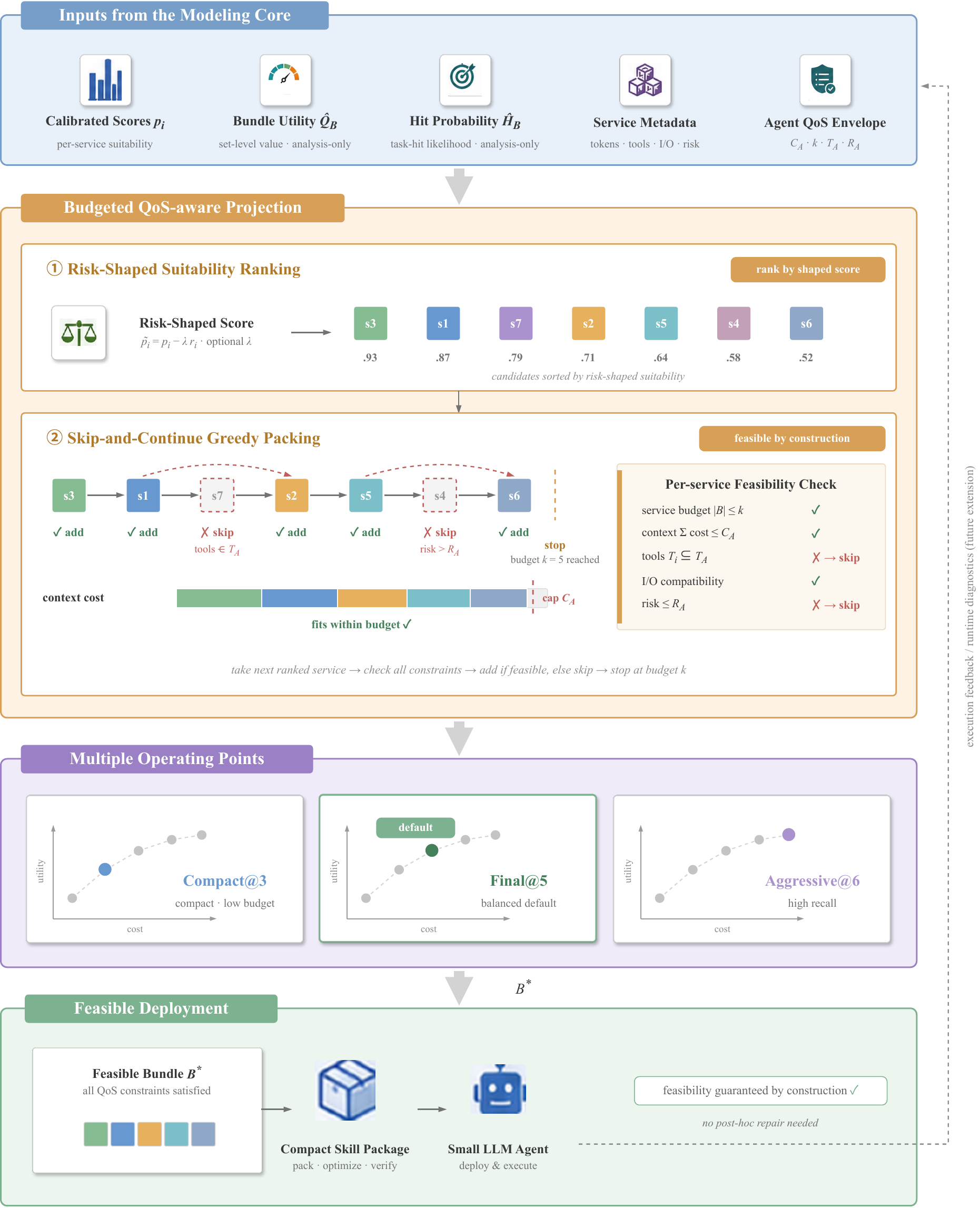}
    \caption{Budgeted QoS-aware projection and deployment output: skip-and-continue packing under token, risk, and tool-availability constraints, followed by optional compact-package adaptation for diagnostic execution. This figure expands Stage~4 (budgeted QoS packing) of the overview figure (Fig.~1) in the main text.}
    \label{fig:budgeted_projection}
\end{figure}

Fig.~\ref{fig:budgeted_projection} details the budgeted QoS-aware projection and the deployment output. The main output is a budgeted skill service bundle rather than a ranked list; the bundle can be used directly as the execution-context input of a small LLM agent, or converted into a task-specific compact skill package by extractive adaptation that keeps key execution steps, tool-use instructions, input-output constraints, and risk notes.

\end{document}